\documentclass[british]{iopart}
\usepackage[T1]{fontenc}
\usepackage[utf8]{inputenc}
\usepackage{color}
\usepackage{babel}
\usepackage{array}
\usepackage{booktabs}
\usepackage{textcomp}
\usepackage{multirow}
\usepackage{amstext}
\usepackage{amssymb}
\usepackage{graphicx}
\usepackage{rotfloat}
\usepackage{esint}
\usepackage{nomencl}
\providecommand{\printnomenclature}{\printglossary}
\providecommand{\makenomenclature}{\makeglossary}
\makenomenclature

\makeatletter

\providecommand{\tabularnewline}{\\}

\usepackage{iopams}
\usepackage{setstack}
\newenvironment{lyxlist}[1]
	{\begin{list}{}
		{\settowidth{\labelwidth}{#1}
		 \setlength{\leftmargin}{\labelwidth}
		 \addtolength{\leftmargin}{\labelsep}
		 }}
	{\end{list}}

\usepackage{etoolbox}
\usepackage{hyperref}

\makeatletter
\def\@mkboth#1#2{}
\newlength\appendixwidth
\preto\appendix{\addtocontents{toc}{\protect\patchl@section}}
\makeatother

\makeatother

\begin{document}
\topical[Negative-$U$ defects in semiconductors]{Characterisation of negative-$U$ defects in semiconductors}
\author{José Coutinho}
\address{i3N, Department of Physics, University of Aveiro, Campus Santiago,
3810-193 Aveiro, Portugal}
\ead{jose.coutinho@ua.pt}
\author{Vladimir P Markevich and Anthony R Peaker}
\address{Photon Science Institute, School of Electrical and Electronic Engineering,
The University of Manchester, Manchester M13 9PL, United Kingdom}
\begin{abstract}
This review aims at providing a retrospective, as well as a description
of the state-of-the-art and future prospects regarding the theoretical
and experimental characterisation of negative-$U$ defects in semiconductors.
This is done by complementing the account with a description of the
work that resulted in some of the most detailed, and yet more complex
defect models in semiconductors. The essential physics underlying
the negative-$U$ behaviour is presented, including electronic correlation,
electron-phonon coupling, disproportionation, defect transition levels
and rates. Techniques for the analysis of the experimental data and
modelling are also introduced, namely defect statistics, kinetics
of carrier capture and emission, defect transformation, configuration
coordinate diagrams and other tools. We finally include a showcase
of several works that led to the identification of some of the most
impacting negative-$U$ defects in group-IV and compound semiconductors.
{[}\emph{Post-print published in Journal of Physics: Condensed Matter
}\textbf{\emph{32}}\emph{, 323001 (2020)}; DOI:\href{https://doi.org/10.1088/1361-648X/ab8091}{10.1088/1361-648X/ab8091}{]}
\end{abstract}

\maketitle
\ioptwocol

\tableofcontents{}

\printnomenclature{}

\nomenclature[U]{$U$}{Electronic correlation energy}\nomenclature[Ueff]{$U_{\mathrm{eff}}$}{Effective correlation energy (includes atomic reorganization energy)}\nomenclature{EPR}{Electron paramagnetic resonance}\nomenclature{DLTS}{Deep level transient spectroscopy}\nomenclature[lambda_ep]{$\lambda_{\mathrm{ep}}$}{Electron-phonon coupling coefficient}\nomenclature[E^q]{$E^{(q)}$}{Total energy of a defect with net charge $q$ with respect to the neutral state}\nomenclature[E^N+n]{$E^{(N+n)}$}{Total energy of a state with $N+n$ electrons, with $N$ being an arbitrary reference.}\nomenclature[E_c]{$E_{\mathrm{c}}$}{Conduction band bottom energy}\nomenclature[E_v]{$E_{\mathrm{v}}$}{Valence band top energy}\nomenclature[E_g]{$E_{\mathrm{g}}$}{Band gap width}\nomenclature[TSCAP]{TSCAP}{Thermally stimulated capacitance}\nomenclature[CC]{CC}{Configuration coordinate (diagram)}\nomenclature[LVM]{LVM}{Local vibrational mode}\nomenclature[OD-ENDOR]{OD-ENDOR}{Optically detected electron-nuclear double resonance}\nomenclature[MPE]{MPE}{Multi-phonon emission}\nomenclature[density of holes]{$p$}{Density of free holes in the valence band top}\nomenclature[density of electrons]{$n$}{Density of free electrons in the conduction band bottom}\nomenclature[$n$-th ionization]{$I_n$}{$n$-th ionization energy with respect to the neutral state}\nomenclature[$n$-th affinity]{$A_n$}{$n$-th electron affinity with respect to the neutral state}\nomenclature[DSCF]{$\Delta$SCF}{Delta self-consistent field}\nomenclature[free hole]{$\mathrm{h}^+$}{Free hole at the top of the valence band}\nomenclature[free electron]{$\mathrm{e}^-$}{Free electron at the bottom of the conduction band}\nomenclature[E_corr]{$E_{\mathrm{corr}}$}{Total electronic correlation energy		}\nomenclature[c_n/p]{$c_{n/p}$}{Capture rate for electrons/holes}\nomenclature[E_F]{$E_{\mathrm{F}}$}{Fermi energy}\nomenclature[E_form]{$E_{\mathrm{f}}$}{Formation energy}\nomenclature[C_n/p]{$C_{n/p}$}{Capture coefficient for electrons/holes}\nomenclature[e_n/p]{$e_{n/p}$}{Emission rate of electrons/holes}\nomenclature[sigma_{n/p}]{$\sigma_{n/p}$}{Apparent capture cross-section for electron/holes}\nomenclature[C_infn/p]{$\sigma_{\infty,n/p}$}{Direct capture cross-section for electron/holes}\nomenclature[v_c/v_th]{$\langle v_{\mathrm{c/v}}\rangle_{\mathrm{th}}$}{Thermal-average velocity of electrons/holes in the conduction/valence band}\nomenclature[N_c/v]{$N_{\mathrm{c/v}}$}{Density of states at the bottom/top of the conduction/valence band}\nomenclature[E(N/N')]{$E(N/N')$}{Transition level of a defect involving a change in the number of bound electrons ($N>N'$)}\nomenclature[E(q/q')]{$E(q/q')$}{Transition level of a defect involving a change in charge state ($q<q'$)}\nomenclature[E_sigma]{$E_{\sigma n/p}$}{Capture barrier for electrons/holes}\nomenclature[R]{$\mathbf{R}$ / $\mathbf{R}_\alpha$}{Collective / individual cartesian coordinates of atoms}\nomenclature[Q]{$\mathbf{Q}$ / $Q_k$}{Generalized coordinate / normal mode of vibration}\nomenclature[N_d]{$N_{\mathrm{d}}$}{Defect concentration}\nomenclature[{[D]}]{[D]}{Concentration of a defect with label/state `D'}\nomenclature[TDD]{TDD}{Thermal double donors}\nomenclature[Cz]{Cz}{Czochralski}\nomenclature[rho]{$\rho$}{Electron density}\nomenclature[f^N]{$f^{(N)}$}{Fraction of defects in a state with $N$ bound electrons}\nomenclature[Infra-red]{IR}{Infra-red}

\cleardoublepage{}

\section{Preface\label{sec:preface}}

The presence of defects and contaminants in crystalline solids, and
in particular crystalline semiconductors, is unavoidable. Defects,
if not deliberately introduced, necessarily occur due to a fundamental
reason: the lowering of free energy by increasing configurational
entropy. Contamination, on the other hand, essentially depends on
the contact of the sample with alien species, and therefore on the
specificities of growth and processing technologies.

Defects and impurities in semiconductors can be intentionally introduced
in order to confer the material with customised mechanical, electrical,
optical or magnetic functionality. However, even when intentionally
present, they may also turn out to be critical as the semiconductor
can acquire other prominent, but unwanted properties. For instance,
while a certain amount of dissolved oxygen is highly beneficial in
terms of the mechanical stability of silicon wafers, under certain
conditions the oxygen impurities precipitate and form clusters which
act as shallow donors and so increase the availability of electrons
for electrical transport at room temperature. This effect alters the
electrical specifications of the silicon, and in p-type material type-inversion
can occur. Another example is the creation of silicon vacancies in
silicon carbide which can act as single-photon emitters for quantum
communications. However, they are also responsible for the introduction
of several deep electron traps which are highly detrimental in terms
of compensation effects in n-type material. Clearly, understanding
defect properties, and if possible, their quantum mechanics, is paramount
for technologies that range from solar power conversion to quantum
computing.

Defects can be classified in many ways, but regarding their geometric
properties they are usually distinguished as point-like and extended.
Examples of point defects are vacancies, interstitials, antisites
(misplaced atoms in compound crystals) and atomic-scale complexes.
On the other hand, extended defects include dislocations, surfaces
and interfaces. Regardless of their size and shape, they change the
potential in the Hamiltonian of the pristine material. Consequently,
that often leads to the appearance of defect states within a spectral
range which would be forbidden to the otherwise perfect crystal.

Unlike crystalline states, defect states are localised in real space.
They may capture charge carriers depending whether the stabilisation
energy (for instance due to bond formation or exchange interactions)
overcomes the repulsive energy between the captured charge and other
charges already present. In some cases, defects may actually capture
more than one carrier of the same type, but now the stabilisation
energy has to exceed the repulsion from previously captured charges
as well. This extra cost is often referred to as correlation energy
($U$). We may find an analogy of this effect in the sequential ionisation
of an atom, for instance,

{\numparts}

\begin{eqnarray}
\textrm{He}^{0}+I_{1}\rightarrow\textrm{He}^{+}+\textrm{e}^{-};\label{CR:Helium1}\\
\textrm{He}^{+}+\textrm{e}^{-}+I_{2}\rightarrow\textrm{He}^{++}+2\textrm{e}^{-},\label{CR:Helium2}
\end{eqnarray}
{\endnumparts}where $I_{n}$ is the $n$-th ionization energy of
the He atom. Here $I_{2}>I_{1}$, and the difference $U=I_{2}-I_{1}$,
which essentially accounts for the electron-electron repulsion and
nuclear screening, is obviously positive. Alternatively, $U$ may
be defined as

\begin{equation}
U=E(\textrm{He}^{0})+E(\textrm{He}^{++})-2E(\textrm{He}^{+}),\label{EQ:Helium3}
\end{equation}
where $E(\textrm{He}^{q})$ is the energy of the He atom in the $q$
charge state. The fact that $U>0$, makes a pair of $\textrm{He}^{+}$
ions more stable than $\textrm{He}^{0}+\textrm{He}^{++}$, so that
a $\textrm{He}^{+}$ gas will not decompose into a mix of $\textrm{He}^{0}$
and $\textrm{He}^{++}$ ensembles.

As in the case of isolated atoms, most defects in semiconductors show
positive $U$ values. However, we know today, that many effectively
show a negative-$U$ correlation between sequential charging events
in apparent defiance of electrostatics. It is as if they became less
`eager' for electrons (holes) immediately after electron (hole)
emission!

In the early 1970's, Anderson introduced the idea of negative-$U$
to explain the absence of paramagnetism in glassy-semiconductors doped
with n- and p-type impurities \cite{Anderson1975}. The material was
envisaged as a random network of three-state bonds whose potential
energy was given by $V=1/2\,c\,x^{2}-\lambda_{\textrm{ep}}(n_{\uparrow}+n_{\downarrow})x+n_{\uparrow}\,n_{\downarrow}\,U$.
Here $x$ represented a bond coordinate, $c$ its respective harmonic
coefficient, and $n_{\uparrow}$ and $n_{\downarrow}$ were spin-up
and spin-down bond occupancies ($n_{\sigma}=\{0,1\}$). The last term
accounted for an electronic Hubbard correlation energy \cite{Hubbard1963},
with impact only when both electrons pair to form the bond ($n_{\uparrow}=n_{\downarrow}=1$).
The striking ingredient of the model was the introduction of an electron-phonon
coupling constant $\lambda_{\textrm{ep}}$ connecting the bond displacement
with its occupancy. Accordingly, considering atomic relaxations and
the definition of Eq.~\ref{EQ:Helium3}, the net effective correlation
involving the three states ($n_{\uparrow}+n_{\downarrow}=\{0,1,2\}$)
is given by $U_{\textrm{eff}}=U-\lambda_{\textrm{ep}}^{2}/c$. It
was then noted that for sufficiently strong coupling, $U_{\textrm{eff}}$
becomes negative, thus providing an explanation for why in many glasses
and polymers, paramagnetic $n_{\uparrow}+n_{\downarrow}=1$ states
are unstable against diamagnetic $n_{\uparrow}+n_{\downarrow}=\{0,2\}$
states.

Although the original idea of Anderson was to describe the formation
of bipolarons (where pairing of two electrons and two holes in a localised
region is favoured against single polaron formation), the analogy
to the negative-$U$ effect in defects is evident. In fact, the concept
was extended to defects in semiconducting glasses (like $\textrm{As}_{2}\textrm{Se}_{3}$)
by Street and Mott \cite{Street1975}. Kastner and co-workers \cite{Kastner1976}
analysed the applicability of the model to several other compounds,
emphasising that the dominant contribution to the negative correlation
energy is of chemical origin, or in other words, is driven by rebonding
of atoms.

In 1979, Baraff, Kane and Schlüter anticipated numerically that neutral,
positive and double-positively charged vacancies in crystalline silicon
($V_{\textrm{Si}}^{0}$, $V_{\textrm{Si}}^{+}$ and $V_{\textrm{Si}}^{++}$,
respectively) formed an `Anderson negative-$U$' system \cite{Baraff1979}.
A Jahn-Teller distortion in the neutral charge state was claimed to
be responsible for the effect, being sufficiently stabilising as to
render $V_{\textrm{Si}}^{+}$ unstable against $V_{\textrm{Si}}^{0}$
and $V_{\textrm{Si}}^{++}$, irrespective of the position of the Fermi
energy. A few months later, these predictions were confirmed experimentally
by Watkins and Troxell \cite{Watkins1980}. The following arguments
provided the grounds for the claim: (i) the paramagnetic state $V_{\textrm{Si}}^{+}$,
as monitored by electron paramagnetic resonance (EPR), was only observed
at cryogenic temperatures under photo-excitation. This was an indication
that $V_{\textrm{Si}}^{+}$ is metastable; (ii) Upon turning off the
light, the intensity of the EPR signal bleached at a rate limited
by a barrier of 0.05~eV. This figure was interpreted as the activation
barrier for hole emission during $V_{\textrm{Si}}^{+}\rightarrow V_{\textrm{Si}}^{0}+\textrm{h}^{+}$;
(iii) From deep level transient spectroscopy (DLTS), a peak with an
activation barrier for hole emission of 0.13~eV, with twice the intensity
expected for a single hole emission, could be connected to the following
two-hole emission sequence,

\begin{equation}
V_{\textrm{Si}}^{++}\rightarrow V_{\textrm{Si}}^{+}+\textrm{h}^{+}\rightarrow V_{\textrm{Si}}^{0}+2\textrm{h}^{+}.\label{CR:VSi1}
\end{equation}
It was then argued that for a negative-$U$ double donor, the hole
involved in the first ionisation was bound more strongly (0.13~eV)
than the second one (0.05~eV). Thus, at a temperature where the first
hole is emitted, the second hole should follow immediately, naturally
accounting for the double intensity of the peak.

Baraff, Watkins and others, inspired an entire community towards the
characterisation of many other negative-$U$ defects in semiconductors.
This account is about their work and the observations and concepts
that followed. It aims at reviewing the literature and presenting
the most recent results regarding experimental and theoretical methods
involved in the characterisation of negative-$U$ defects in semiconductors.

The main text provides a survey regarding several experimental and
theoretical reports on negative-$U$ defects (Section~\ref{sec:intro}),
an introduction to the physical concepts involved in the description
of negative-$U$ defects (Section~\ref{sec:negu-defects}), details
regarding experimental and theoretical methods that are used for their
characterisation (Section~\ref{sec:characterisation}), a tabular
showcase of a selection of negative-$U$ defects in semiconductors,
accompanied by a detailed description of some of the most interesting
and technologically relevant ones (Section~\ref{sec:showcase}),
and finally a revision of the main concepts, results and challenges
for the future of the topic (Section~\ref{sec:conclusions}).

\section{Introduction\label{sec:intro}}

Multi-stability is among the most fascinating properties of point
defects in semiconductors \cite{Benton1989}. In the present context,
we refer to a multi-stable defect as one which can be found in several
inequivalent atomic configurations for a particular charge state.
Along the same lines, a bi-stable defect has two non-degenerate atomic
structures in the same charge state. The relative stability as well
as formation/annihilation rates of different configurations of multi-stable
defects are often sensitive to the application of external stimuli
(\emph{e.g.} temperature, electro-magnetic fields or mechanical stress).
Hence, under favourable conditions, their populations can be shifted
from those observed under equilibrium.

Upon changing the atomistic structure of defects, a change of its
electronic structure also takes place. It is therefore expected that
stimuli-induced defect reconfigurations may affect significantly the
properties of the host material. Exhibition of persistent photoconductivity,
photo-induced capacitance quenching, or temperature-dependent carrier
trapping are symptoms that are commonly connected to multi-stability.
The magnitude and timescale of these effects depend not only on the
properties of the defects themselves, like transformation barriers,
cross-sections and transition dipole moments, but also on sample properties
like defect concentration and distribution, its thermal history, and
of course the measurement conditions.

\begin{figure}
\begin{centering}
\includegraphics[width=7.5cm]{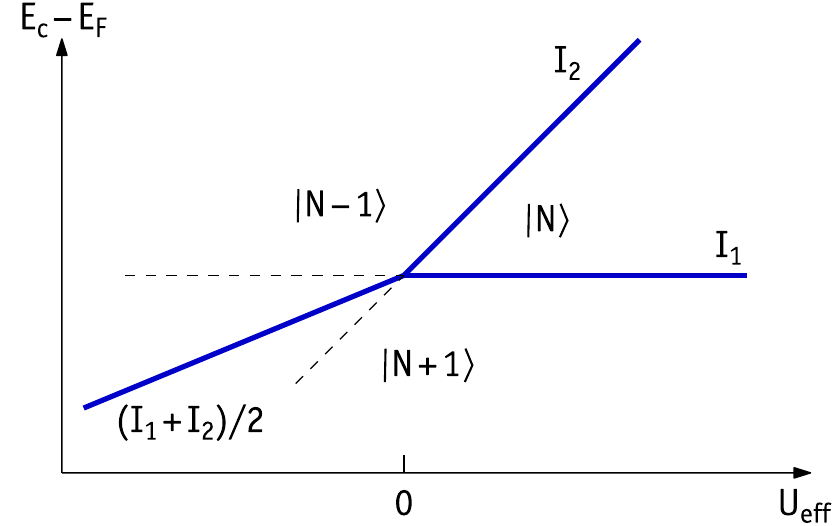}
\par\end{centering}
\caption{\label{FIG:PHASEDIA}Phase diagram of a three-charge-state defect,
with electron occupancy $|N\!-\!1\rangle$, $|N\rangle$, and $|N\!+\!1\rangle$.
The horizontal axis is the effective correlation energy ($U_{\textrm{eff}}=I_{2}-I_{1}$),
whereas the vertical axis is the Fermi energy with respect to the
conduction band bottom ($E_{\textrm{c}}-E_{\textrm{F}}$). First and
second ionisation energies of the defect are $I_{1}$ and $I_{2}$,
respectively. Phase boundaries (blue lines) define the stability borders
$(U_{\textrm{eff}},E_{\textrm{F}})$ between different charge states.
(Adapted from Ref.~\cite{Langer1983}).}
\end{figure}

Negative-$U$ defects are intimately tied to multi-stability and metastability.
These defects have at least three charge states, say with electron
occupancy $|N\!-\!1\rangle$, $|N\rangle$, and $|N\!+\!1\rangle$.
Here $N$ is an arbitrary reference and the $|N\!-\!m\rangle$ state
also includes $m$ electrons at a reservoir with energy $E_{\textrm{F}}$
per electron (Fermi level). The distinction of negative-$U$ defects
is that the intermediate $|N\rangle$ state is metastable and not
found under equilibrium conditions. Figure~\ref{FIG:PHASEDIA} shows
a schematic phase diagram of such a defect, where transition boundaries
are drawn as a function of the effective correlation energy ($U_{\textrm{eff}}=I_{2}-I_{1}$)
and $E_{\textrm{F}}$. Here $I_{m}=E^{(N+1-m)}-E^{(N+2-m)}$ is the
$m$-th ionization energy, with $E^{(N)}$ standing for the energy
of state $|N\rangle$. When $U_{\textrm{eff}}>0$, all three charge
states can be populated under equilibrium conditions, depending on
the location of the Fermi energy with respect to the transition levels
at $E_{\textrm{c}}-I_{1}$ and $E_{\textrm{c}}-I_{2}$, where $E_{\textrm{c}}$
is the conduction band bottom. However, for a negative-$U$ defect,
only $|N\!-\!1\rangle$ or $|N\!+\!1\rangle$ states are observed,
with equal populations being found when the Fermi level is located
at $E_{\textrm{c}}-(I_{1}+I_{2})/2$.

Negative-$U$ defects show strong lattice relaxation effects upon
capture and emission of carriers. For that reason, charge state transitions
may involve considerable barriers and large Franck-Condon shifts,
often making optical absorption and luminescence data hardly comparable
to measurements carried out at equilibrium conditions (\emph{e.g.}
Hall effect). The metastable $|N\rangle$ state of negative-$U$ defects
has a short lifetime, being formed only when sample conditions are
away from equilibrium. Although reactions like $|N\!-\!1\rangle\rightarrow|N\rangle+\textrm{h}^{+}$
or $|N\!+\!1\rangle\rightarrow|N\rangle+\textrm{e}^{-}$ can in principle
be stimulated with the application of external perturbations, \emph{e.g.}
injection of current or light pulsing, the population of $|N\rangle$
critically depends on a delicate balance between the time of the measurement,
the forward and backward reaction (decay) rates, the latter being
thermodynamically favoured.

Another difficulty for detection of negative-$U$ defects is related
to the fact that the stability of $|N\!-\!1\rangle$ and $|N\!+\!1\rangle$
states is usually connected to the formation of closed-shell diamagnetic
states, and therefore they are undetectable by EPR \cite{Anderson1975}.
These are among the many features which make the characterisation
of negative-$U$ defects a challenge, not only in the laboratory,
but also from the perspective of theory and modelling.

The decade of the 1980's was particularly prolific regarding the design
and conception of experiments capable of probing the properties of
negative-$U$ defects. Watkins and co-workers showed that in DLTS
experiments, after applying filling pulses to Si diodes irradiated
with electrons, carriers trapped at single vacancies ($V_{\textrm{Si}}$)
and boron interstitials ($\textrm{B}_{\textrm{i}}$) in Si, were subsequently
emitted in pairs under reverse bias \cite{Watkins1980,Troxell1980}.
This is a particular signature of the negative-$U$ property, which
follows from the relatively fast rate of the second emission compared
to the slow rate of the first one. To large extent, this is determined
by the condition $I_{1}>I_{2}$. Hence, a conventional DLTS experiment
can only measure a single transient, with a decay characteristic of
the first (slower) emission. For the case of $\textrm{B}_{\textrm{i}}$,
the measured transient corresponded to electron emission from the
negatively charged $\textrm{B}_{\textrm{i}}^{-}$ (the deeper $|N\!+\!1\rangle$
state) \cite{Troxell1980}. Direct detection of the shallower emission
$\textrm{B}_{\textrm{i}}^{0}\rightarrow\textrm{B}_{\textrm{i}}^{+}+\textrm{e}^{-}$
(from the $|N\rangle$ state) was not possible. The series of electrical
trap-filling pulses required for DLTS measurements, quickly filled
all traps into the $|N\!+\!1\rangle$ state, no matter how short the
applied pulses were. This was later solved by replacing the biased
injection of electrons by optical injection into boron-doped diodes
with a mesa structure \cite{Harris1982,Harris1987}.

A prominent example of a multi-stable defect which shows a negative-$U$
ordering of electronic transitions is the M-centre in InP. This defect
is observed in electron irradiated (1~MeV) undoped, nominally n-type
InP, and was found in at least two main forms (A and B), depending
on the bias conditions of the sample \cite{Levinson1983a,Levinson1983b,Stavola1984}.
A clear indication that the A-form shows a negative-$U$ ordering
of transition levels was found from thermally stimulated capacitance
(TSCAP) measurements. While most transitions indicated an ordinary
heating-induced loss of a single charge accumulated in a trap, two
of them (labelled A2 and A3), occurred simultaneously, thus pointing
to a very specific feature of negative-$U$ defects: a two-electron
emission event. Despite the detailed experimental data already reported
for the M-centre, the only atomistic model available in the literature
is phenomenological. It consists of an indium-vacancy-phosphorous-antisite
($V_{\textrm{In}}\textrm{P}_{\textrm{In}}$) and phosphorous-vacancy-phosphorous-antisite-pair
($V_{\textrm{P}}$--$2\textrm{P}_{\textrm{In}}$), which were assigned
to A and B forms, respectively \cite{Wager1985}.

Another defect showcasing the negative-$U$ effect was firstly reported
by Henry and Lang in order to explain the observation of persistent
photoconductivity in chalcogen-doped $\textrm{Al}_{x}\textrm{Ga}_{1-x}\textrm{As}$
alloys \cite{Henry1977,Lang1979}. There was no doubt that the defect
involved a shallow donor species (D). However, because it also had
a deep state at $\mbox{\ensuremath{\sim\!E_{\textrm{c}}-0.1}}$~eV,
postulated at the time as due to complexing with an undetermined constituent
(X), the defect was labelled `DX' centre. Based on first-principles
pseudopotential calculations, Chadi and Chang came up with a model
of a deep DX acceptor state localised on a broken $\textrm{S}_{\textrm{As}}\textrm{-Ga}$
(or Al) bond \cite{Chadi1988}. Accordingly, when the Ga (Al) atom
connected to neutral $\textrm{S}_{\textrm{As}}$ (referred to as $d^{0}$)
moves away along the $\langle111\rangle$ crystallographic axis towards
the interstitial site, the total energy increases as expected for
a stable structure, but the donor state becomes progressively deeper.
When the level has dropped by more than the local repulsive correlation,
the total energy is now $E_{\textrm{cap}}\sim0.4$~eV above the minimum
energy of $d^{0}$. At this point a free electron (donated by another
$\textrm{S}_{\textrm{As}}$ dopant in the sample) is captured. The
subsequent atomistic relaxation stabilises the negative state by $E_{\textrm{e}}=-0.5$~eV,
with $\textrm{DX}^{-}$ landing at $E_{\textrm{cap}}-E_{\textrm{e}}=0.1$~eV
below the $d^{0}$ state. The authors underlined that the disproportionation
reaction $2d^{0}\rightarrow d^{+}+\textrm{DX}^{-}$ was exothermal
and therefore indicative of a negative-$U$ defect. This provided
a natural explanation for the lack of EPR involving the $\textrm{DX}^{-}$
state.

A refinement of the model was subsequently proposed by Dobaczewski
and Kaczor \cite{Dobaczewski1991}, who carried out a detailed analysis
of the photoionisation of DX in Te-doped $\textrm{Al}_{x}\textrm{Ga}_{1-x}\textrm{As}$.
In their work, the intermediate metastable state could better describe
the ionisation kinetics if assumed to be a localised $\textrm{DX}^{0}$
defect (as opposed to the neutral shallow donor proposed by Chadi
and Chang \cite{Chadi1988}). It is however consensual that the observed
persistent photoconductivity can be accounted for as resulting from
the optical ionisation of an equilibrated population of $\textrm{DX}^{-}$
(in the $|N\!+\!1\rangle$ state) into metastable $\textrm{DX}^{0}+\textrm{e}^{-}$
($|N\rangle$ state), which quickly converts into $d^{+}+2\textrm{e}^{-}$
($|N\!-\!1\rangle$ state). The latter persists due to a large capture
barrier hindering the recovery of $\textrm{DX}^{-}$.

Many other reports of negative-$U$ defects followed the above examples,
including defects with huge impact on the electronic properties of
semiconductors and devices. Prominent examples are the early members
of the thermal double donor family of defects in Si and Ge \cite{Latushko1986,Litvinov1988},
interstitial hydrogen which can be involved in a multitude of solid-state
reactions with defects and dopants in several semiconductors \cite{Pankove1991,Pearton1992},
the carbon vacancy in 4H- and 6H-SiC which decisively limits the life-time
of minority carriers in n-type material \cite{Hemmingsson1998,Hemmingsson1999},
or boron-oxygen complexes involved in the light induced degradation
of solar Si \cite{VaqueiroContreras2019}.

The characterisation of negative-$U$ defects involves the experimental
monitoring of a two-carrier emission/capture reaction $|N\!+\!1\rangle\rightleftarrows|N\rangle+\textrm{e}^{-}\rightleftarrows|N\!-\!1\rangle+2\textrm{e}^{-}$.
This is a multi-step process involving both electronic transitions
and geometric transformations. The kinetics is usually limited by
a slow step, effectively `masking' any subsequent fast steps. Under
these circumstances, special tricks have been introduced in order
to access the individual steps experimentally. Examples are the combination
of junction spectroscopy with light excitation, or the judicious control
of the amount of free carriers available for capture by using very
short and limited injection pulses. From the above, it is also clear
that theory, in particular first-principles modelling of the electronic
structure, has played a huge role in unveiling the workings of negative-$U$
defects and providing guidance for experiments. It is in this spirit
of collaboration that we intend to introduce the reader to the topic
of negative-$U$ defects in semiconductors. Besides presenting a survey
regarding what has been achieved so far, we dedicated much of the
content to the concepts and techniques (both experimental and theoretical)
involved in the characterisation of this class of defects. We end
up with the identification of several problems, from minor loose ends
to completely obscure issues, in the hope of stirring up those who
may feel challenged by the topic.

\section{Negative-$U$ defects in semiconductors\label{sec:negu-defects}}

\subsection{Electronic correlation\label{subsec:correlation}}

Electronic correlation can be thought of as a measure of how much
entangled two or more electrons are, or alternatively, how hard electron
motion is within the field of other electrons. It is instructive to
look at the concept from the perspective of the Hartree-Fock (HF)
method \cite{Levine2014}. Accordingly, the wave function is represented
by a Slater determinant of $M$ spin-orbitals $\phi_{i}(\mathbf{x})=\psi_{i}(\mathbf{r})\sigma(s_{i})$
with space- and spin-dependence $\psi_{i}(\mathbf{r})$ and $s_{i}=\{\uparrow,\downarrow\}$,
respectively, where $i=1,\ldots,M$ is a state-index. The manifold
$\mathbf{x}$ is a composite of space and spin coordinates. The total
energy has contributions from single-electron and two-electron components
\cite{Blinder2019},

\begin{equation}
E_{\textrm{HF}}=\sum_{i}H_{i}+\sum_{i>j}\left(J_{ij}-K_{ij}\right),\label{EQ:EHF}
\end{equation}
where \emph{core integrals},

\begin{equation}
H_{i}=\int\textrm{d}\mathbf{r}^{3}\,\psi_{i}^{*}(\mathbf{r})\left(-\frac{1}{2}\nabla_{i}^{2}-\sum_{\alpha}\frac{Z_{\alpha}}{\left|\mathbf{r}-\mathbf{R}_{\alpha}\right|}\right)\psi_{i}(\mathbf{r})\label{EQ:HFCORE}
\end{equation}
describe a set of independent electrons orbiting in the field of nuclei
located at coordinates $\mathbf{R}_{\alpha}$. Here, the Born-Oppenheimer
approximation holds, so that $\mathbf{R}_{\alpha}$ are fixed (the
nuclear kinetic energy vanishes, whereas the nuclear-nuclear repulsion
terms add up to an obvious constant which can be included in the result
after computing the electronic energy). Interactions between electrons,
namely their mutual repulsion and exchange contributions are given
by the \emph{Coulomb} and \emph{exchange} integrals,

\begin{equation}
J_{ij}=\iint\textrm{d}\mathbf{r}^{3}\textrm{d}\mathbf{r}'^{3}\,\frac{\left|\psi_{i}(\mathbf{r})\right|^{2}\left|\psi_{j}(\mathbf{r}')\right|^{2}}{\left|\mathbf{r}-\mathbf{r}'\right|}\label{EQ:HFCOUL}
\end{equation}
and

\begin{equation}
K_{ij}=\iint\textrm{d}\mathbf{r}^{3}\textrm{d}\mathbf{r}'^{3}\,\psi_{i}^{*}(\mathbf{r})\psi_{j}^{*}(\mathbf{r}')\frac{\langle\sigma(s_{i})|\sigma(s_{j})\rangle}{\left|\mathbf{r}-\mathbf{r}'\right|}\psi_{i}(\mathbf{r}')\psi_{j}(\mathbf{r}),\label{EQ:HFEXX}
\end{equation}
respectively, and they obey to the relation $J_{ij}\geq K_{ij}\geq0$.
Unlike $J_{ij}$ elements, exchange terms involve the interaction
between states with parallel spin only. The scalar product $\langle\sigma(s_{i})|\sigma(s_{j})\rangle=0$
when $s_{i}\neq s_{j}$, \emph{i.e.} when the interacting spin-orbitals
have opposite spin. This follows from the anti-symmetry of the wave
function with respect to the exchange of any electronic coordinate.

Equations~\ref{EQ:HFCOUL} and \ref{EQ:HFEXX} translate that each
electron effectively interacts with an average charge distribution
due to other electrons. This mean-field approximation ignores dynamical
correlation, \emph{i.e.}, that electron coordinates are inexorably
interdependent, so that they instantaneously avoid one another to
reduce their mutual repulsion as much as possible. Nondynamical correlation
is also disregarded since the wave function is approximated by a single
Slater determinant. Both effects were referred to by Löwdin \cite{Lowdin1955}
as total correlation energy, $E_{\textrm{corr}}=E-E_{\textrm{HF}}$
, where $E$ is the exact non-relativistic energy of the Schrödinger
equation. $E_{\textrm{corr}}$ is a very difficult quantity to obtain
-- a prohibitively large number (millions) of Slater determinants
are needed in order to obtain a wave function approaching the exact
solution, thus making the problem intractable, even for today’s largest
supercomputers \cite{Szabo1996}.

If we are solely interested in the ground state, significant progress
can be made with the use of density functional theory (DFT) \cite{Hohenberg1964,Parr1989,Burke2012},
which replaces the HF total energy (a functional of the wave function)
by a functional of the electron density $\rho$ \cite{Kohn1965},

\begin{eqnarray}
E[\rho(\mathbf{r})] & = & T_{0}[\rho(\mathbf{r})]+\int\textrm{d}\mathbf{r}^{3}\,v_{\textrm{ext}}(\mathbf{r})\rho(\mathbf{r})+\nonumber \\
 & + & \frac{1}{2}\iint\textrm{d}\mathbf{r}^{3}\textrm{d}\mathbf{r}'^{3}\,\frac{\rho(\mathbf{r})\rho(\mathbf{r}')}{\left|\mathbf{r}-\mathbf{r}'\right|}+E_{\textrm{xc}}[\rho(\mathbf{r})].\label{EQ:EDFT}
\end{eqnarray}
Analogous to the core integrals $H_{i}$ of Eq.~\ref{EQ:EHF}, the
energy functional accounts for the kinetic energy of an arbitrary
number of non-interacting electrons ($T_{0}$), subject to an external
potential ($v_{\textrm{ext}}$), due to nuclear and other non-electronic
fields. The electron density is obtained from summation over occupied
orbitals, $\rho(\mathbf{r})=\sum_{i=\textrm{occ}}|\psi_{i}(\mathbf{r})|^{2}$
, which satisfy a set of partial differential equations analogous
to single-particle Schrödinger equations. These are solved self-consistently
in order to obtain the ground state density (which depends solely
on the external potential) \cite{Kohn1965}. The electron-electron
Coulomb repulsion term in Eq.~\ref{EQ:EDFT}) is analogous to the
contribution of $J_{ij}$ integrals in HF, although it now includes
an unphysical self-interaction energy (electrons are repelled by themselves).
In a similar fashion to the exchange integrals $K_{ij}$, the exchange
and correlation functional $E_{\textrm{xc}}$ describes the electronic
exchange interactions, but also incorporates all remaining effects,
including correlation (absent in HF) and the neutralisation of the
above-mentioned spurious self-interactions. Unfortunately, the mathematical
form of $E_{\textrm{xc}}$ is not known (except for an homogeneous
electron gas \cite{Dirac1930,Parr1989,Chachiyo2016}). However, several
approximations have been proposed, varying in physical detail and
accuracy, and of course in computational load (see for instance Ref.~\cite{Burke2012}
and references therein).

A very simple and intuitive method of treating correlation is that
proposed by Hubbard for narrow electronic bands \cite{Hubbard1963}.
It nicely describes electronic motion within systems whose electrons
are strongly localised on atomic orbitals like Mott insulators. Within
the second-quantisation formalism, the Hubbard Hamiltonian is given
by \cite{Lieb2003},

\begin{equation}
\hat{H}_{\textrm{Hubb}}=t\sum_{\langle ij\rangle,\sigma}\hat{c}_{i,\sigma}^{\dagger}\hat{c}_{i,\sigma}+U\sum_{i}\hat{n}_{i,\uparrow}\hat{n}_{i,\downarrow}\label{EQ:HUB}
\end{equation}
where $\langle ij\rangle$ means that the summation runs over nearest-neighbour
atomic sites $i$ and $j$, while $\hat{c}_{i,\sigma}^{\dagger}$,
$\hat{c}_{i,\sigma}$ and $\hat{n}_{i,\sigma}$ stand for creation,
annihilation and number operators for electrons of spin $\sigma$
on site $i$. For strongly localised states, electronic motion proceeds
via hopping at a rate proportional to the jump integral $t$, which
relates to the band width. This is described by the first term of
Eq.~\ref{EQ:HUB} and involves electron transfer between neighbouring
sites $\langle ij\rangle$. The second term accounts for the fact
that each site is capable of accommodating up to two electrons of
opposite spin. However, when a site is fully occupied ($\hat{n}_{i,\uparrow}=\hat{n}_{i,\downarrow}=1$),
both electrons interact with an energy $U$. Hence, according to the
Hubbard model, correlation is the energy raise during an electron
transfer event involving two neighbouring sites with identical occupancy
(one electron each).

\begin{figure}
\begin{centering}
\includegraphics[width=7.5cm]{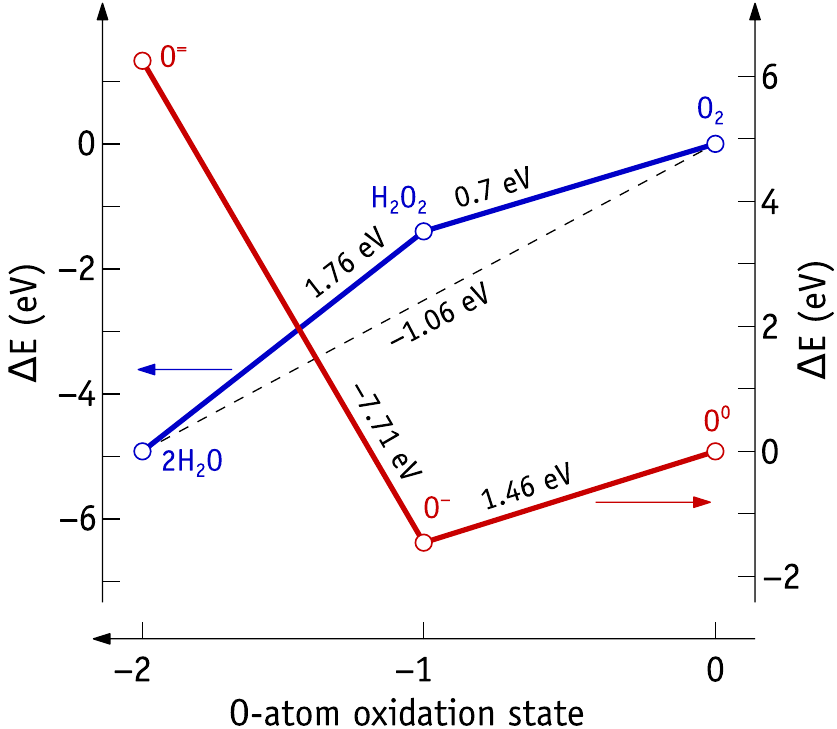}
\par\end{centering}
\caption{\label{FIG:FROST}Frost diagram showing the first and second electron
affinities of atomic oxygen (red) \cite{Chaibi2010}, along with disproportionation
energetics of hydrogen peroxide ($\textrm{H}_{2}\textrm{O}_{2}$)
in aqueous solution (blue) \cite{Shaffer2017}. Numbers edging solid
and dashed line segments represent first- and second-order variations,
respectively, of the free-energy per oxygen atom upon oxidation.}
\end{figure}

Let us look at the above concept using a pair of oxygen atoms as an
example. We know that the energy of a generic atomic system is a piecewise
linear function of the electron occupancy (see for instance Refs.~\cite{Perdew1982}).
This is illustrated by the Frost diagram of Figure~\ref{FIG:FROST}
(red line), which shows the change in the total energy for the sequential
reduction of atomic oxygen in contact with an electron reservoir (read
the diagram from right to left along the horizontal axis). Oxygen
is a highly electronegative species with positive first electron affinity
($A_{1}=1.46$~eV), and because of that, $\textrm{O}^{-}$ is more
stable than the neutral atom. However, capture of an additional electron
is not favourable due to accumulated Coulomb repulsion. This is exhibited
by a negative second affinity, $A_{2}=-7.71$~eV. From the diagram,
it is also clear that any (even fractional) charge transfer $0<\delta\leq1$
between a pair of $\textrm{O}^{-}$ anions, $2\textrm{O}^{-}\rightarrow\textrm{O}^{-1-\delta}+\textrm{O}^{-1+\delta}$,
raises the total energy by a correlation energy $\Delta E^{(N+\delta)}=(A_{1}-A_{2})\delta$.
For the transfer of a whole electron ($\delta=1$), we obtain a general
expression for the total correlation of a reference state with $N$
electrons.

\begin{equation}
U^{(N)}=\Delta^{(N+\delta)}=E^{(N+1)}+E^{(N-1)}-2E^{(N)},\label{EQ:CORR1}
\end{equation}
where the energies are indexed to the total number of electrons on
each state. For the particular case of the $\textrm{O}^{-}$ atomic
state, the correlation energy is $U=1.46+7.71=9.17$~eV.

Strictly speaking, correlation is exclusively an electronic property,
and that is unambiguously defined for a single atom by Eq.~\ref{EQ:CORR1}.
However, for a more complex moiety (\emph{e.g.} a molecule, atom in
solution, defect in a solid, \emph{etc.}), the atomistic geometry,
vibrational modes and electronic structure of ionised $|N\!+\!1\rangle$
and $|N\!-\!1\rangle$ states may differ substantially from the those
of the reference $|N\rangle$-electron state. It is convenient to
recall at this point the concept of effective correlation energy,
which besides the electronic contribution, includes a relaxation energy,

\begin{eqnarray}
U_{\textrm{eff}}^{(N)} & = & U^{(N)}-\left(\Delta U_{\textrm{rel}}^{(N+1)}+\Delta U_{\textrm{rel}}^{(N-1)}\right)\nonumber \\
 & = & U^{(N)}-\Delta U_{\textrm{rel}}^{(N)}.\label{EQ:CORR2}
\end{eqnarray}
In the above, $U^{(N)}$ is the electronic correlation as defined
at the geometry of the $N$-electron reference state. On the other
hand, $\Delta U_{\textrm{rel}}^{(N+1)}$ and $\Delta U_{\textrm{rel}}^{(N-1)}$
are positive relaxation energies, respectively affecting the $|N\!+\!1\rangle$
and $|N\!-\!1\rangle$ states just after transitioning from the N-electron
state. These quantities reflect ionisation-induced atomic reconfigurations,
entropy changes, \emph{etc.}, and they are analogous to the Franck-Condon
energy dissipated by a molecule or defect following absorption or
luminescent transitions \cite{Franck1926}.

\subsection{Disproportionation\label{subsec:disproportionation}}

An electron transfer event involving two identical moieties with $N$
electrons can be broken into two steps: (i) a sudden electronic excitation
where an electron-hole pair is created -- this raises the energy
by $U^{(N)}$, and (ii) a subsequent electronic/atomistic relaxation
that lowers the energy by $\Delta U_{\textrm{rel}}^{(N)}$. According
to Eqs.~\ref{EQ:CORR1} and \ref{EQ:CORR2}, for a relaxation as
large as $\Delta U_{\textrm{rel}}^{(N)}>U^{(N)}$, we have $2E^{(N)}>E^{(N+1)}+E^{(N-1)}$.
In that case, a close pair of $N$-electron moieties becomes metastable
and disproportionates into $N-1$ and $N+1$ electron species. Disproportionation
is a well-known effect in electrochemistry, either taking place spontaneously
or via thermal activation, eventually with the help of a catalyst
\cite{Wilson1948}. Although disproportionation necessarily involves
electron transfer, it is essentially driven by atomic rearrangement.
An eloquent example is given by the reaction

\begin{equation}
2\textrm{Cu}^{+}\,(\textrm{aq})\rightarrow\textrm{Cu}\,(\textrm{s})+\textrm{Cu}^{++}\,(\textrm{aq})\label{CR:Cu}
\end{equation}
where univalent copper (in solution) precipitates into solid Cu and
cupric ions. Being exothermic, Reaction~\ref{CR:Cu} gives $2E(\textrm{Cu}^{+})>E(\textrm{Cu}^{0})+E(\textrm{Cu}^{++})$,
with $E(\textrm{Cu}^{q})$ being the free energy per Cu species in
the oxidation state $q$ in its respective phase. In this reaction,
the relaxation energy $\Delta U_{\textrm{rel}}$ essentially results
from the formation of Cu bonds in the copper metal.

Another textbook example of a disproportionation reaction is the decomposition
of hydrogen peroxide accelerated at a metallic surface \cite{Kremer1959},

\begin{equation}
2\textrm{H}_{2}\textrm{O}_{2}^{-}\,(\textrm{aq})\stackrel{\textrm{cat}}{\longrightarrow}2\textrm{H}_{2}\textrm{O}^{=}\,(\textrm{l})+\textrm{O}_{2}\,(\textrm{g}).\label{CR:H2O2}
\end{equation}
The superscripts in the componds above stand for the oxidation state
of oxygen. According to the reaction, for every $\textrm{O}^{-}$-pair
in $\textrm{H}_{2}\textrm{O}_{2}$, one oxygen atom is reduced (to
form water), while another becomes oxidised (leading to formation
of molecular oxygen). The Frost diagram for Reaction~\ref{CR:H2O2}
is represented in Figure~\ref{FIG:FROST} (blue line). Using Eq.~\ref{EQ:CORR1},
we find that the correlation energy per oxygen atom in $\textrm{H}_{2}\textrm{O}_{2}$
is

\begin{eqnarray}
U_{\textrm{eff}} & = & \left[E(2\textrm{H}_{2}\textrm{O})+E(\textrm{O}_{2})-2E(\textrm{H}_{2}\textrm{O}_{2})\right]/2\label{EQ:CORR4}\\
 & = & -1.06\,\textrm{eV}
\end{eqnarray}
The factor of $1/2$ arises due to the fact that each compound has
two oxygen atoms. The negative effective correlation energy tells
us that $\textrm{O}^{-}$ ions in hydrogen peroxide are metastable
against disproportionation into $\textrm{O}^{=}$ an $\textrm{O}^{0}$
in water and molecular oxygen, respectively. In Figure~\ref{FIG:FROST},
$U_{\textrm{eff}}$ is the free energy difference per oxygen atom,
between the $\textrm{H}_{2}\textrm{O}_{2}\,(\textrm{aq})$ state and
a mix of $\textrm{H}_{2}\textrm{O}\,(\textrm{l})$ and $\textrm{O}_{2}\,(\textrm{g})$
represented by the dashed line. Like in the copper ion redox Reaction~\ref{CR:Cu},
the effective correlation of $\textrm{O}^{-}$ ions in $\textrm{H}_{2}\textrm{O}_{2}$
is negative due to atomic reconfiguration, more specifically due to
proton transfer and formation of oxygen double bonds after electron
transfer.

Many other examples are found in the literature, including the solid
state disproportionation of tin(II) oxide (SnO) into tin(VI) and tin
dioxide (SnO$_{2}$), again involving dramatic structural transformations
\cite{Giefers2005}, or the loss of bond length translational order
in several perovskites due to random disproportionation of the metallic
ions in the lattice \cite{Dalpian2018}.

\subsection{Electronic transition levels of defects\label{subsec:levels}}

The term \emph{electronic transition level} of a defect has been given
different meanings depending on the context. It has been interpreted
differently, depending whether one refers to single-particle or many-body
calculations, if it embodies electron-phonon coupling or not, or perhaps
if it relates to measurements based on thermodynamic or kinetic quantities.
In the present case, an electronic transition level of a defect (or
simply \emph{defect level}), is denoted as \cite{Shockley1957},

\begin{equation}
E(N\!\!+\!\!1\,/\,N)=E^{(N+1)}-E^{(N)},\label{EQ:LEVEL1}
\end{equation}
and corresponds to the Fermi energy of a material, $E_{\textrm{F}}$,
above which the ground state of $N+1$ electrons bound to a defect
is more stable than any $N$-electron state (plus one electron at
the Fermi reservoir). Analogously, one could say that when $E_{\textrm{F}}$
drops below a $E(N\!\!+\!\!1\,/\,N)$ level, the $(N\!+\!1)$-electron
state becomes unstable against the $N$-electron state (plus one electron
at the Fermi reservoir). When the Fermi level coincides with the defect
level, $N$- and $(N\!+\!1)$-state populations are identical. Under
such conditions, the following isothermic conversion between $\textrm{D}^{(N)}$
and $\textrm{D}^{(N+1)}$,

\begin{equation}
D^{(N)}+\textrm{e}^{-}\rightleftarrows D^{(N+1)},\label{CR:DEF1}
\end{equation}
shows identical forward and backward rates and occurs via exchange
of electrons ($\textrm{e}^{-}$) with a reservoir with Fermi energy
$E_{\textrm{F}}=E(N\!\!+\!\!1\,/\,N)$.

Let us assume that we have a concentration $[\textrm{D}]$ of non-interacting
and identical defects in a sample at temperature $T$. We also postulate
that for the allowed range of $E_{\textrm{F}}$ values, the defects
may occur in up to two charge states, with respective concentrations
$[\textrm{D}^{(N)}]$ and $[\textrm{D}^{(N+1)}]$, thus possessing
a single electronic transition level $E(N\!\!+\!\!1\,/\,N)$. Under
equilibrium, the fraction of defects with $N$ bound electrons is
given by \cite{Shockley1957,Look1981,Sah1996},

\begin{equation}
f^{(N)}=Z^{-1}g^{(N)}\exp\left(-E_{\textrm{f}}^{(N)}/k_{\textrm{B}}T\right),\label{EQ:BOLTZ}
\end{equation}
where $k_{\textrm{B}}$ is the Boltzmann constant, $Z$ the partition
function,

\begin{equation}
Z=\sum_{m}g^{(m)}\exp\left(-E_{\textrm{f}}^{(m)}/k_{\textrm{B}}T\right),\label{EQ:PART}
\end{equation}
and $g^{(m)}$ is a degeneracy factor \cite{Look1981}. In Eqs.~\ref{EQ:BOLTZ}
and \ref{EQ:PART}, $E_{\textrm{f}}^{(m)}$ is the formation energy
of a stable $m$-electron state,

\begin{equation}
E_{\textrm{f}}^{(m)}=E^{(m)}-mE_{\textrm{F}}\label{EQ:EFRM1}
\end{equation}
with the second term on the right of Eq.~\ref{EQ:EFRM1} representing
the chemical potential of electrons in a reservoir with Fermi energy
$E_{\textrm{F}}$. For our two-state defect this results in the well-known
distribution function,

\begin{equation}
f^{(N+1)}=\left[1+\left(\frac{g^{(N)}}{g^{(N+1)}}\right)\exp\left(\frac{E(N\!\!+\!\!1\,/\,N)-E_{\textrm{F}}}{k_{\textrm{B}}T}\right)\right]^{-1}.\label{EQ:FRAC1}
\end{equation}
Hence, if $E_{\textrm{F}}$ is lowered below $E(N\!\!+\!\!1\,/\,N)$,
the backward rate of Reaction~\ref{CR:DEF1} exceeds the forward
rate, and most defects bind $N$ electrons. Conversely, if $E_{\textrm{F}}$
is raised above $E(N\!\!+\!\!1\,/\,N)$, occupation of the next available
bound state becomes energetically favourable and the reaction proceeds
to the right. Defined in this way, an electronic level depends on
ground state energies $E^{(N)}$ and $E^{(N+1)}$ only. It is a thermodynamic
quantity, analogous to a critical chemical potential for which a phase
transition takes place.

\begin{figure}
\begin{centering}
\includegraphics[width=7.5cm]{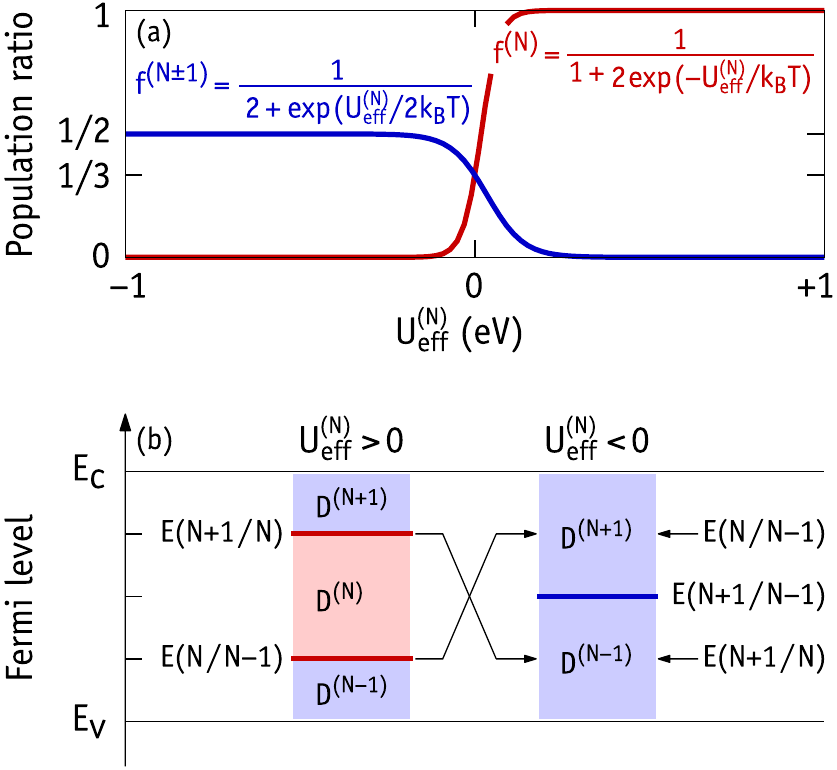}
\par\end{centering}
\caption{\label{FIG:POPULATION}(a) Population fraction of $|N\rangle$ and
$|N\pm1\rangle$ defect states as a function of the effective correlation
$U_{\textrm{eff}}^{(N)}=E(N\!\!+\!\!1\,/\,N)-E(N\,/\,N\!\!-\!\!1)$.
The Fermi level is located at half-way between $E(N\!\!+\!\!1\,/\,N)$
and $E(N\,/\,N\!\!-\!\!1)$ transition levels. Other assumptions are
a band gap width of $E_{\textrm{g}}=1$~eV, $T=300$~K, and $g^{(N-1)}=g^{(N)}=g^{(N+1)}=1$.
(b) Transition level diagram for positive- and negative-$U$ defects
(left and right, respectively). Transition levels are shown as horizontal
thick lines. Labels in the shaded areas indicate the most abundant
(stable) defect state under equilibrium conditions for the different
values of $E_{\textrm{F}}$ .}
\end{figure}

Next we consider the case of a defect with two transition levels in
the band gap. These define the borders between three charge states,
say $\textrm{D}^{(N-1)}$, $\textrm{D}^{(N)}$ and $\textrm{D}^{(N+1)}$,
in phase space. Let us inspect the populations of the three charge
states by varying the level positions, but keeping the Fermi level
locked at midway between $E(N\,/\,N\!\!-\!\!1)$ and $E(N\!\!+\!\!1\,/\,N)$,
\emph{i.e.}, $E_{\textrm{F}}=[E^{(N+1)}-E^{(N-1)}]/2$. Using Eq.~\ref{EQ:BOLTZ},
we readily arrive at,

\begin{equation}
f^{(N)}=\left[1+\frac{g^{(N+1)}+g^{(N-1)}}{g^{(N)}}\exp\left(-\frac{U_{\textrm{eff}}^{(N)}}{k_{\textrm{B}}T}\right)\right]^{-1}\label{EQ:POPN}
\end{equation}

\begin{equation}
f^{(N\pm1)}=\left[1+\frac{g^{(N\mp1)}}{g^{(N\pm1)}}+\frac{g^{(N)}}{g^{(N\pm1)}}\exp\left(\frac{U_{\textrm{eff}}^{(N)}}{2k_{\textrm{B}}T}\right)\right]^{-1}\label{EQ:POPN+-1}
\end{equation}
with

\begin{equation}
U_{\textrm{eff}}^{(N)}=E(N\!\!+\!\!1\,/\,N)-E(N\,/\,N\!\!-\!\!1).\label{EQ:CORR5}
\end{equation}

The above fractions are plotted in Figure~\ref{FIG:POPULATION}(a),
where we assume that $g^{(N-1)}=g^{(N)}=g^{(N+1)}=1$. Under these
conditions $f^{(N+1)}=f^{(N-1)}$, thus being represented by a common
function $f^{(N\pm1)}$. The above degeneracy factors do not influence
the location of the transition levels when extrapolated to $T\rightarrow0$.
In the graph of Figure~\ref{FIG:POPULATION}(a) we also assume that
the band gap width is $E_{\textrm{g}}=1$~eV and $T=300$~K.

For a positive-$U$ defect ($U_{\textrm{eff}}^{(N)}>0$), we have
$f^{(N)}\approx1$ (see right-hand side of Figure~\ref{FIG:POPULATION}(a))
and the population of the other two states is negligible. On the other
hand, for a negative-$U$ defect ($U_{\textrm{eff}}^{(N)}<0$), $\textrm{D}^{(N)}$
is metastable as seen by the negligible probability of finding this
state (left-hand side of Figure~\ref{FIG:POPULATION}(a)). Under
these conditions, the reaction

\begin{equation}
2\textrm{D}^{(N)}\rightleftarrows\textrm{D}^{(N+1)}+\textrm{D}^{(N-1)},\label{CR:DEF3}
\end{equation}
becomes an exothermic disproportionation process with an energy balance,

\begin{equation}
\Delta E_{\textrm{r}}=E(N\!\!+\!\!1\,/\,N)-E(N\,/\,N\!\!-\!\!1)=U_{\textrm{eff}}^{(N)}.\label{EQ:DER}
\end{equation}

When the Fermi level is in the middle of the two levels with negative-$U$
ordering, $\textrm{D}^{(N-1)}$ and $\textrm{D}^{(N+1)}$ show identical
populations $f^{(N\pm1)}=1/2$, implying the existence of a $E(N\!\!+\!\!1\,/\,N\!\!-\!\!1)$
transition level at

\begin{equation}
E(N\!\!+\!\!1\,/\,N\!\!-\!\!1)=[E(N\!\!+\!\!1\,/\,N)+E(N\,/\,N\!\!-\!\!1)]/2,\label{EQ:LEVNU}
\end{equation}
which is represented by the thick blue line in the right-hand side
of Figure~\ref{FIG:POPULATION}(b). The $E(N\,/\,N\!\!-\!\!1)$ and
$\mbox{\ensuremath{E(N\!\!+\!\!1\,/\,N)}}$ levels involve the metastable
$\textrm{D}^{(N)}$ state, so that they cannot correspond to thermodynamic
transition levels. Their location is indicated by the arrows in Figure~\ref{FIG:POPULATION}(b).
Instead, they determine the position of the thermodynamic $\mbox{\ensuremath{E(N\!+\!1\,\,/\,\,N\!-\!1)}}$
transition, which involves the exchange of two electrons with the
Fermi reservoir. Finally, for the peculiar case of $U_{\textrm{eff}}^{(N)}=0$,
Figure~\ref{FIG:POPULATION}(a) indicates that $f^{(N)}=f^{(N\pm1)}=1/3$
so that a triple-point with all states equally populated is attained.
Under these conditions, all states in Reaction~\ref{CR:DEF3} are
equilibrated under isothermic conditions ($\Delta E_{\textrm{r}}=0$).

\begin{figure*}
\begin{centering}
\includegraphics[width=11cm]{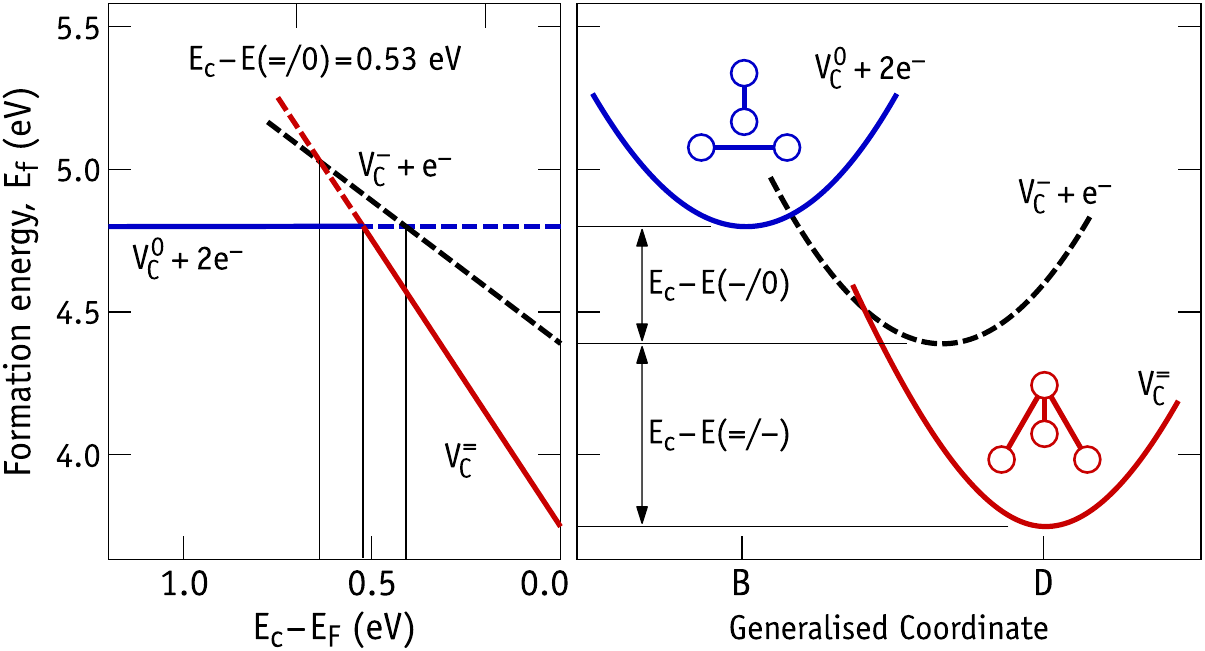}
\par\end{centering}
\caption{\label{FIG:CC-VC-SIC}Formation energy diagram (left) along with a
configuration coordinate diagram (right) for the carbon vacancy at
the pseudo-cubic site in n-type 4H-SiC. Structures B and D are represented
as four Si atoms (white circles) forming an approximate tetrahedron
viewed along the main crystallographic axis. The vacant site lies
below the Si atom at the centre. Atoms connected by segments are separated
by a shorter distance in comparison to equivalent atoms in bulk \cite{Coutinho2017}.}
\end{figure*}

\subsection{Formation energy diagrams\label{subsec:formdiagram}}

In the dilute limit, where defect-defect interactions are negligible,
the equilibrium concentration of a defect on a charge state $q$ depends
on its formation energy $E_{\textrm{f}}^{(q)}$ as,

\begin{equation}
[\textrm{D}^{(q)}]=[\textrm{D}_{0}^{(q)}]\exp(-E_{\textrm{f}}^{(q)}/k_{\textrm{B}}T),\label{EQ:DCON}
\end{equation}
with $[\textrm{D}_{0}^{(q)}]$ being the density of degenerate configurations
(lattice sites and orientations per unit volume) that the defect has
in the sample. In the above, we now use the charge state $q$ to label
the electronic state of the defect (instead of the number of electrons
or holes). The formation energy in Eq.~\ref{EQ:DCON} expresses the
cost to create an isolated defect by trading electrons and atomic
species with electronic and atomic reservoirs with chemical potentials
$E_{\textrm{F}}$ and $\mu_{i}$, respectively (see Ref.~\cite{Todorova2014}
and references therein),

\begin{equation}
E_{\textrm{f}}^{(q)}(\mu_{i},E_{\textrm{F}})=E^{(q)}-\sum_{i}n_{i}\mu_{i}+qE_{\textrm{F}},\label{EQ:EFRM2}
\end{equation}
were $E^{(q)}$ is the energy per defect in charge state $q$, surrounded
by a sufficiently large volume of host material, enclosing $n_{i}$
atoms of species $i$ (dilute limit). We note that in Eq.~\ref{EQ:DCON}
we did not consider the contribution of entropy to the formation energy.
To do so, $E_{\textrm{f}}^{(q)}$ would have to be replaced by a free
energy of formation, and $E^{(q)}$ in Eq.~\ref{EQ:EFRM2} would
become $H^{(q)}-TS^{(q)}$, with $H^{(q)}$ and $S^{(q)}$ being the
enthalpy and entropy per defect in charge state $q$, the later accounting
for vibrational, electronic and magnetic entropy terms. While the
formation entropy can be of significant importance, in particular
at high temperatures \cite{Grabowski2009}, for defects with deep
states in the gap, electronic excitations depend exponentially on
large activation energies, so that electronic and magnetic entropy
can often be neglected. More care has to be taken regarding the vibrational
entropy, particularly for negative-$U$ defects where distinct structures
with rather different vibrational spectra can occur. Although being
usually small when compared to the electronic transitions of deep
defects, vibrational entropy can be easily incorporated in Eq.~\ref{EQ:EFRM2}
(see for instance Estreicher et~al. \cite{Estreicher2004}).

Defect formation energies are in principle positive quantities, otherwise
the host material would become unstable against spontaneous defect
creation. The graphical representation of the formation energy of
a defect as a function of the Fermi level (for a fixed set of atomic
chemical potentials) is a common procedure in defect physics and chemistry.
It allows us to compare the relative binding energy of electrons to
different defects (with variable charge and stoichiometry) with respect
to a common reference -- the Fermi level.

On the left-hand side of Figure~\ref{FIG:CC-VC-SIC} we depict a
formation energy diagram of the carbon vacancy ($V_{\textrm{C}}$)
located on the pseudo-cubic site of 4H-SiC, for $E_{\textrm{F}}$
values in the upper part of the band gap (n-type material). The diagram
was constructed solely based on experimental observations. The origin
of the horizontal axis (right limit of the axis) corresponds to the
conduction band bottom. As we move to the left, $E_{\textrm{F}}$
lowers within the band gap.

The $V_{\textrm{C}}$ defect at the pseudo-cubic site is a double
acceptor that shows negative-$U$ ordering of levels. These were measured
by conventional DLTS and Laplace-DLTS at $E_{\textrm{c}}-E(=/-)=0.64$~eV
and $E_{\textrm{c}}-E(-/0)=0.41$~eV \cite{Hemmingsson1998,Capan2018}.
From Eq.~\ref{EQ:LEVNU} and taking the singly negative state as
reference, the acceptor states show an effective negative correlation
energy of $U_{\textrm{eff}}=-0.23$~eV, and that implies the existence
of a thermodynamic transition level at,

\begin{equation}
E(=/0)=[E(=/-)+E(-/0)]/2=E_{\textrm{c}}-0.53\,\textrm{eV},\label{EQ:LEVSIC}
\end{equation}
while $E(=/-)$ and $E(-/0)$ are metastable. The $E(=/0)$ level
establishes that under equilibrium conditions, only $V_{\textrm{C}}^{0}$
or $V_{\textrm{C}}^{=}$ states can be found in n-type 4H-SiC. This
is highlighted in Figure~\ref{FIG:CC-VC-SIC} as solid and dashed
lines for the representation of the formation energy of stable and
metastable states, respectively. That explains why observation of
the $V_{\textrm{C}}^{-}$ paramagnetic state requires optical excitation
\cite{Son2012}. All three relevant transitions are indicated in the
horizontal axis of Figure~\ref{FIG:CC-VC-SIC} by vertical lines
at the crossing points of the formation energy segments. The slope
of each segment is $qE_{\textrm{F}}$, clearly distinguishing the
charge state of the respective $V_{\textrm{C}}^{q}$ defect. The charge
neutralisation condition holds via trading of $(2-q)\textrm{e}^{-}$
electrons with the Fermi reservoir. The last piece of information
needed to construct the formation energy diagram of Figure~\ref{FIG:CC-VC-SIC}
is the formation energy of the neutral species ($V_{\textrm{C}}^{0}$).
This was measured by means of high temperature annealing/quenching
experiments as $E_{\textrm{f}}^{(0)}=4.8$~eV \cite{Ayedh2015}.

\subsection{Configuration coordinate diagrams\label{subsec:ccdiagram}}

While formation energy diagrams capture defect thermodynamics, they
are not so useful when it comes to representing the kinetics of transitions
which are actually measured by many techniques, including optical
spectroscopy and DLTS. In essence, a typical experiment involves monitoring
the response of a sample subjected to the application/removal of an
external perturbation (e.g. applied field or temperature), leading
to a displacement/recovery of the equilibrium conditions. The measured
excitation/relaxation rates of defects in semiconductors may involve
absorption/emission of light, the exchange of energy and momentum
with phonons and electrons within the sample, or a combination of
all these processes. Although being applicable to any type of electronic
transitions, we introduce the concept of a configuration coordinate
(CC) diagram to illustrate the vibronic nature of non-radiative capture
and thermal emission of carriers, which due to the conservation of
energy, involve a trade of phonons with the host lattice \cite{Henry1977,Stoneham1981}.
This is the most common type of transitions for negative-$U$ defects,
where strong electron-phonon coupling effects are normally found.

\begin{figure}
\begin{centering}
\includegraphics[width=7.5cm]{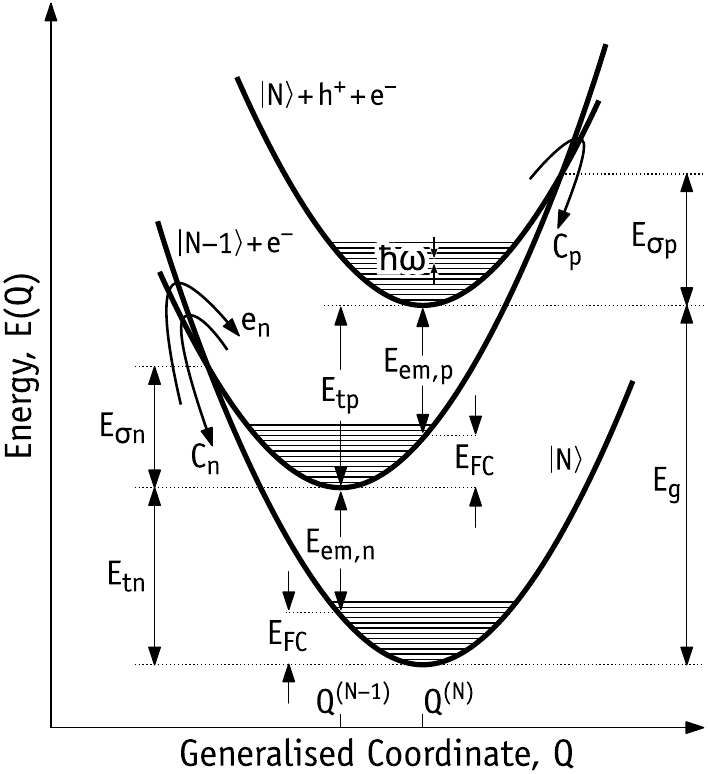}
\par\end{centering}
\caption{\label{FIG:CC-DIA}Configuration coordinate diagram of a defect with
a transition level responsible for an electron trap located at $E_{\textrm{c}}-E(N\,/\,N\!\!-\!\!1)$.
Each parabola represents a vibronic state that depends on a generalised
coordinate of atoms. Closely spaced horizontal segments represent
effective phonon quanta of energy. At the bottom, we show the $|N\rangle$
state, which can emit an electron and become $|N-1\rangle+\textrm{e}^{-}$,
where $\textrm{e}^{-}$ is a free electron at the conduction band
bottom. At the top, we show the $|N\rangle$ state plus an uncorrelated
electron-hole pair. $C_{n/p}$ are capture coefficients for electrons/holes,
whereas $e_{n}$ are emission rates for electrons. Electron/hole capture
barriers are $E_{\sigma n/p}$. (Reprinted from Ref.~\cite{Peaker2018},
with the permission of AIP Publishing).}
\end{figure}

The process of non-radiative phonon assisted transitions is depicted
in the CC diagram of Figure~\ref{FIG:CC-DIA}, where we assume that
the whole system has a single effective vibrational degree of freedom.
The $|N\rangle$ ground state finds its minimum at the generalised
atomic coordinate $Q^{(N)}$. The upper parabola, which is displaced
up in energy by the band gap width ($E_{\textrm{g}}$), represents
an excited state comprising a $|N\rangle$ defect plus an uncorrelated
electron-hole pair (eventually created after illumination of the sample
with above-bandgap light). In the middle of the CC diagram we represent
a $|N-1\rangle$ state, which due to electron-phonon coupling, has
a minimum-energy coordinate shifted to $Q^{(N-1)}$. To avoid unnecessary
complications, we assume that the curvature of both $|N\rangle$ and
$|N-1\rangle$ parabola corresponds to the same vibrational frequency
with quanta $\hbar\omega$.

Thermal activated conversion of $|N\rangle$ into $|N-1\rangle$ can
be achieved in two ways: either (i) upon capture of a free-hole followed
by multi-phonon emission, or (ii) via emission of a bound electron
into the conduction band assisted by multi-phonon capture \cite{Peaker2018}.
In the first case the hole capture coefficient is given by

\begin{equation}
C_{p}=\sigma_{p}\langle v_{\textrm{v}}\rangle_{\textrm{th}}p,\label{EQ:CAP1}
\end{equation}
where $\sigma_{p}$ is the apparent capture cross-section for holes
traveling in the valence band top with thermal-average velocity $\langle v_{\textrm{v}}\rangle_{\textrm{th}}$
and density $p$. In the second case, the emission rate is

\begin{equation}
e_{n}=\sigma_{n}\langle v_{\textrm{c}}\rangle_{\textrm{th}}N_{\textrm{c}}\frac{g^{(N-1)}}{g^{(N)}}\exp\left(\frac{E_{c}-E(N\,/\,N\!\!-\!\!1)}{k_{\textrm{B}}T}\right),\label{EQ:EMI1}
\end{equation}
with $N_{\textrm{c}}$ being the density of available states in the
conduction band and $g^{(N-1)}/g^{(N)}$ the degeneracy ratio of the
final to the initial state. The quantities $\sigma_{n}$ and $\langle v_{\textrm{c}}\rangle_{\textrm{th}}$
are electron-analogues of $\sigma_{p}$ and $\langle v_{\textrm{v}}\rangle_{\textrm{th}}$.

It is important to note that the above transitions occur close to
the crossing points of the energy curves. Henry and Lang \cite{Henry1977}
have shown that the uncertainty principle implies that the adiabatic
approximation (upon which the wave function is instantly consistent
with the potential of the oscillating nuclei), breaks down close to
these points, so that transitions may in fact occur for $Q$ values
in a range where $|E^{(N)}-E^{(N-1)}|\lesssim60$~meV. This feature
is incorporated in the temperature dependence of the effective capture
cross-section,

\begin{equation}
\sigma=\sigma_{\infty}\exp(-E_{\sigma}/k_{\textrm{B}}T),\label{EQ:CCSEC}
\end{equation}
with $\sigma_{\infty}$ being the geometrical ($T$-independent) capture
cross section. Capture rates are therefore thermally activated (see
Section~\ref{subsubsec:capture} for further details), and according
to Figure~\ref{FIG:CC-DIA}, the capture barrier for electrons or
holes is

\begin{equation}
E_{\sigma n/p}=\frac{(E_{\textrm{t}n/p}-E_{\textrm{FC}})^{2}}{E_{\textrm{FC}}},\label{EQ:CAPBAR}
\end{equation}
where the depth of the trap for electrons is

\begin{equation}
E_{\textrm{t}n}=E_{\textrm{c}}-E(N\,/\,N\!\!-\!\!1)=E_{\textrm{em},n}+E_{\textrm{FC}},\label{EQ:ETRAPN}
\end{equation}
or for holes,

\begin{equation}
E_{\textrm{t}p}=E(N\,/\,N\!\!-\!\!1)-E_{\textrm{v}}=E_{\textrm{em},p}+E_{\textrm{FC}}.\label{EQ:ETRAPP}
\end{equation}
In the above, trap energies are divided into two components: a \emph{vertical
emission} and a \emph{Franck-Condon relaxation}, $E_{\textrm{em},n/p}$
and $E_{\textrm{FC}}$, respectively (see Figure~\ref{FIG:CC-DIA}).
Within a single-mode approximation, the latter relates to the vibrational
frequency by the Huang-Rhys factor as $E_{\textrm{FC}}=S_{\textrm{HR}}\hbar\omega$,
which for strongly coupled transitions discloses the dissipation of
many phonons ($S_{\textrm{HR}}\gg1$) \cite{Stoneham2001}.

We can now use our example of the carbon vacancy (double acceptor)
in 4H-SiC, and combine its CC and formation energy diagrams to obtain
a consistent and insightful picture of the measurements. The early
DLTS experiments by Hemmingsson \emph{et~al.} \cite{Hemmingsson1998}
have found that the conspicuous Z$_{1/2}$ peak corresponds to the
superposition of two nearly identical Z$_{1}$ and Z$_{2}$ negative-$U$
defects, differing only on the sub-lattice location. These measurements
were recently refined by Koizumi \emph{et~al.} \cite{Koizumi2013}
in 6H-SiC and by Capan \emph{et~al.} \cite{Capan2018} in 4H-SiC,
using Laplace-DLTS. The negative-$U$ ordering of levels implies that
during filling of the defect, the binding energy of the second electron
is higher than that of the first one. On the other hand, for the reverse
process, the thermal emission of the first electron immediately induces
a second emission.

The above is better perceived with help of Figure~\ref{FIG:CC-VC-SIC}.
Accordingly, before pulsing in the majority carriers (n-type material),
the diode is under reverse bias and the Fermi level is lower than
the acceptor levels. The formation energy diagram of Figure~\ref{FIG:CC-VC-SIC}
indicates that the stable state is the neutral one ($V_{\textrm{C}}^{0}$).
After applying a filling pulse (zero or forward bias), the Fermi level
edges the conduction band minimum and the CC diagram on the right-hand
side of Figure~\ref{FIG:CC-VC-SIC} applies. In these conditions
the double negative state ($V_{\textrm{C}}^{=}$) is more stable and
the kinetics of the reaction $V_{\textrm{C}}^{0}+2\textrm{e}^{-}\rightarrow V_{\textrm{C}}^{=}$
is essentially limited by the second capture event \cite{Hemmingsson1998,Capan2018}
-- the negative charge accumulated on the defect after the first
capture makes it less effective for a second capture. Interestingly,
what makes $V_{\textrm{C}}$ a negative-$U$ defect are strong pseudo-Jahn-Teller
relaxations that mainly affect the close shell states (neutral and
double negative) \cite{Coutinho2017}, driving $V_{\textrm{C}}^{-}$
to become metastable (see dashed black lines in Figure~\ref{FIG:CC-VC-SIC}).
The relevant structures (labelled B and D), which were found by first-principles
calculations \cite{Coutinho2017}, are indicated in the CC diagram,
where atoms connected by segments show considerably shorter distances
than those in bulk SiC.

When reverse bias is restored, $V_{\textrm{C}}^{=}$ becomes unstable,
and if the temperature is high enough, two electrons are emitted into
the conduction band bottom and swept away from the depletion region.
We note that the electron capture barriers for $V_{\textrm{C}}^{0}$
and $V_{\textrm{C}}^{-}$ (at the pseudo-cubic site) were found to
be as low as $E_{\sigma n}^{(0)}\sim0$~eV and $E_{\sigma n}^{(-)}\sim0.03$~eV,
respectively \cite{Capan2018}. Considering the relative depth of
both acceptor levels with respect to $E_{\textrm{c}}$, electron emission
from $V_{\textrm{C}}^{=}$ has a higher barrier than that from $V_{\textrm{C}}^{-}$
so that $e_{n}^{(=)}<e_{n}^{(-)}$ (where $e_{n}^{(q)}$ is the electron
emission rate from $V_{\textrm{C}}^{q}$). Hence, conventional DLTS
is only able to monitor a transient whose decay reflects the first
(slower) emission only. This feature is clearly shown in Figure~\ref{FIG:CC-VC-SIC},
illustrating a major difficulty regarding the characterisation of
negative-$U$ defects -- probing the intermediate metastable state.

In the particular case of $V_{\textrm{C}}$ the metastable state is
the singly negative defect. Access to this state was achieved by optical
excitation \cite{Son2012} and keeping the sample at low temperatures
to avoid carrier re-emission, or providing a small amount of electrons
to a reverse-biased n-type diode, via short injection \cite{Koizumi2013,Capan2018}
or optical pulses \cite{Hemmingsson1998}. The injection level must
be well below saturation limit so that the fraction of defects in
the $V_{\textrm{C}}^{=}$ state is much smaller than that in the $V_{\textrm{C}}^{-}$
state.

Although we can only observe a single emission peak by means of junction
spectroscopy, the underlying double emission sequence leads to a variation
of the capacitance twice as large to that observed for a transition
involving a single emission. As we will see in the next Section, this
stands as a rather characteristic feature which is often helpful in
the identification of negative-$U$ defects.

\section{Methods of characterisation\label{sec:characterisation}}

\subsection{Probing negative-$U$ defects\label{subsec:probing}}

For the sake of convenience, below we refer to the effective correlation
energy simply as $U\equiv U_{\textrm{eff}}$. It has already been
mentioned above that under equilibrium conditions, defects with negative-$U$
properties emit or capture charge carriers by pairs. So, a confirmation
of the negative-$U$ property of a defect requires evidence of such
paired emission or capture of charge carriers.

The first definitive experimental evidence of negative-$U$ properties
of a defect in a crystalline semiconductor, the lattice vacancy in
silicon ($V_{\textrm{Si}}$), was obtained from DLTS measurements
\cite{Watkins1980}. This technique is frequently used for the determination
of the concentration of defects with deep levels. Accordingly, the
magnitude of a measured signal (change in sample capacitance due to
carrier emission from a defect), is usually directly proportional
to the concentration the defect trap it refers to, $\Delta C\sim[\textrm{D}]$
\cite{Lang1974}. However, it was found in Ref.~\cite{Watkins1980}
that the magnitude of the DLTS signal due to hole emission from $V_{\textrm{Si}}^{++}$
was proportional to $2[V_{\textrm{Si}}]$, so confirming that each
emission event was in fact a two-hole emission sequence, so the vacancy
in Si is a defect with $U<0$.

It should be mentioned that for the conclusion about $\Delta C\sim2[\textrm{D}]$
in the recorded DLTS spectra, an independent method of determination
of the vacancy concentration was used. Reliable independent methods
for the determination of the concentration of a defect in semiconductor
samples for DLTS measurements are rarely possible, so from a conventional
analysis of emission signals in the DLTS spectra, it is usually not
possible to judge if a carrier emission signal is related to a defect
with positive or negative effective correlation energy. Simple analysis
of temperature dependencies of carrier emission rates measured with
DLTS for a negative-$U$ defect gives only information about parameters
for the emission (activation energy for emission and apparent capture
cross section) of the first, more strongly bound, charge carrier.

\subsubsection{Elucidation of the nature of a defect by means of analysis of its
occupancy with charge carriers at equilibrium conditions\label{subsubsec:elucid-negu}}

Solid evidence of negative-$U$ properties for a number of defects
in semiconductors has been obtained from studies of their occupancy
with charge carriers at equilibrium conditions. The general statistics
of charge distribution at equilibrium conditions for defects with
several trapping levels in semiconductors has been presented in 1957
by Shockley and Last \cite{Shockley1957}. It has been shown that
the concentration ratio of defects in two charge states differing
by one electron is

\begin{equation}
\frac{[\textrm{D}^{(N)}]}{[\textrm{D}^{(N-1)}]}=\exp\left(-\frac{E(N\,/\,N\!\!-\!\!1)-E_{\textrm{F}}}{k_{\textrm{B}}T}\right),\label{EQ:EXP1}
\end{equation}
where $N$ is the number of electrons in the most negative state and
$E(N\,/\,N\!\!-1)$ is the energy level of the defect (\emph{c.f.}
Section~\ref{subsec:levels}).

For a defect with two energy levels, the relationship between them
can by represented by

\begin{equation}
E(N\!\!+\!\!1\,/\,N)=E(N\,/\,N\!\!-\!\!1)+U.\label{EQ:EXP2}
\end{equation}
If $U<0$ and $|U|\gg k_{\textrm{B}}T$, it follows from Eq.~\ref{EQ:EXP1}
that for any position of $E_{\textrm{F}}$, $[\textrm{D}^{(N)}]\ll[\textrm{D}^{(N+1)}]+[\textrm{D}^{(N-1)}]$,
\emph{i.e.} $[\textrm{D}^{(N+1)}]+[\textrm{D}^{(N-1)}]\cong[\textrm{D}]$,
where $[\textrm{D}]$ is the total concentration of the defect. Furthermore,
the electron occupancy function of the $|N+1\rangle$ state is

\begin{eqnarray}
f_{U<0}^{(N+1)} & = & \frac{[D^{(N+1)}]}{[D]}=\nonumber \\
 & = & \left\{ 1+\exp\left[2\frac{E(N\!\!+\!\!1\,/\,N\!\!-\!\!1)-E_{\textrm{F}}}{k_{\textrm{B}}T}\right]\right\} ^{-1},\label{EQ:EXP3}
\end{eqnarray}
where the quantity $E(N\!\!+\!\!1\,/\,N\!\!-\!\!1)=[E(N\!\!+\!\!1\,/\,N)+E(N\,/\,N\!\!-\!\!1)]/2$
represents the occupancy level of the defect with $U<0$. When $E_{\textrm{F}}>E(N\!\!+\!\!1\,/\,N\!\!-\!\!1)$,
the defect has $N+1$ electrons, when $E_{\textrm{F}}<E(N\!\!+\!\!1\,/\,N\!\!-\!\!1)$,
the defect has $N-1$ electrons. It is the difference between Equation~\ref{EQ:EXP3}
and the Fermi function, the later describing the occupancy of single-electron
defect levels, that allows us to determine the nature of a defect
(the sign of the effective electron correlation energy) from a study
of its occupancy with charge carriers.

\begin{figure}
\begin{centering}
\includegraphics[width=7.5cm]{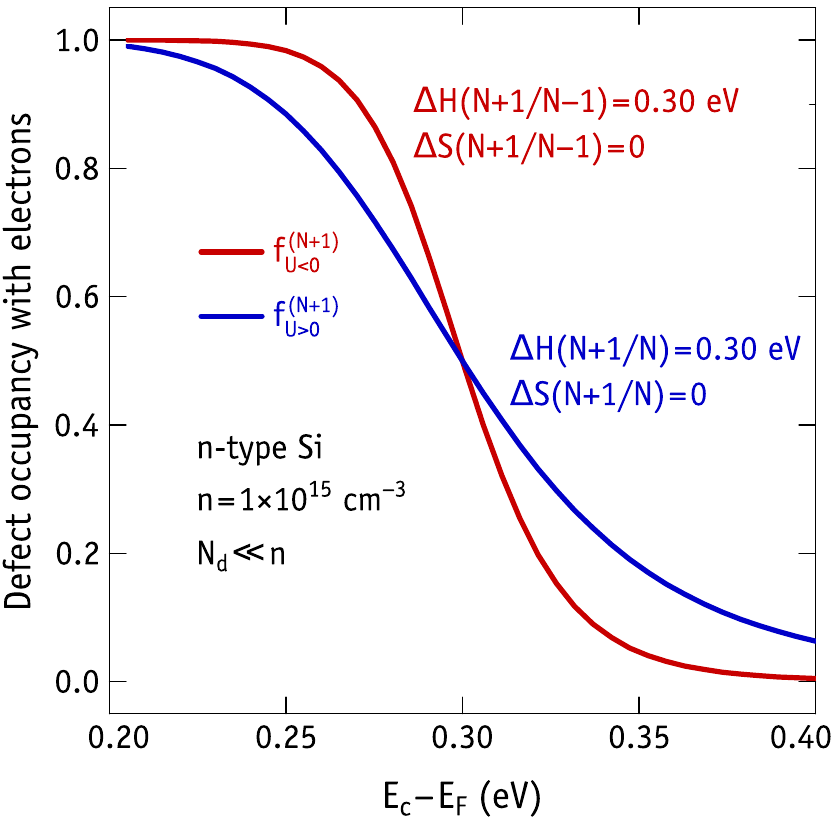}
\par\end{centering}
\caption{\label{FIG:OCCUPATION}Dependencies of occupancy with electrons ($f^{(N+1)}$)
versus the Fermi level position relative to the conduction band edge
for a defect with $U<0$ and $E_{\textrm{c}}-E(N\!\!+\!\!1\,/\,N\!\!-\!\!1)=0.3$~eV
and for a defect with $U>0$ and $E_{\textrm{c}}-E(N\!\!+\!\!1\,/\,N)=0.3$~eV.
The dependencies have been calculated with the use of Eq.~\ref{EQ:EXP3}
(for $U<0$) and the Fermi function (for $U>0$) upon the assumption
$n=1\times10^{15}\,\textrm{cm}^{-3}\gg[\textrm{D}]$.}
\end{figure}

The free energy of ionisation with respect to a reference level (for
instance the conduction band bottom), can be presented as

\begin{eqnarray}
\Delta E(N\!\!+\!\!1\,/\,N\!\!-\!\!1) & = & E_{\textrm{c}}-E(N\!\!+\!\!1\,/\,N\!\!-\!\!1)\label{EQ:FEI1}\\
 & = & \Delta H-T\Delta S,\label{EQ:FEI2}
\end{eqnarray}
where $\Delta H$ and $\Delta S$ are the changes in enthalpy and
entropy due to the $(N\!+\!1\,/\,N\!-\!1)$ ionisation event, and
it is temperature dependent if $\Delta S(N\!+\!1\,/\,N\!-\!1)\neq0$.
As an illustration, Figure~\ref{FIG:OCCUPATION} compares dependencies
of charge occupancy versus Fermi level position in the gap (i) for
a defect with $U<0$ and $E_{\textrm{c}}-E(N\!\!+\!\!1\,/\,N\!\!-\!\!1)=0.3$~eV
and (ii) for a defect with $U>0$ and $E_{\textrm{c}}-E(N\!\!+\!\!1\,/\,N)=0.3$~eV
in n-type silicon, with fully ionised shallow donors providing a density
$n=1\times10^{15}\,\textrm{cm}^{-3}$ of free electrons. The dependencies
have been calculated assuming that $n\gg[\textrm{D}]$, and changes
in the Fermi level position have been induced by temperature variations.

The simplest way to probe the occupancy of a defect with charge carriers
is to change the temperature of a semiconductor sample in a certain
range and monitor associated changes either in free carrier concentration
measured by means of Hall effect, or in magnitude of a signal related
to emission or capture of charge carriers measured by means of DLTS.
Negative-$U$ properties for a number of defects in semiconductors
have been elucidated from analyses of temperature dependencies of
free carrier concentrations, $n\,(p)\sim f(T)$. Particularly, evidence
of negative-$U$ properties of the Si vacancy \cite{Mukashev1982,Emtsev1983},
oxygen-related thermal double donors (TDDs) in silicon and germanium
\cite{Makarenko1985,Latushko1986,Litvinov1988}, and complexes consisting
of a Si interstitial atom with oxygen dimer ($I$O$_{2}$) \cite{Markevich2005a}
and interstitial carbon, oxygen and hydrogen atoms ($\textrm{C}_{\textrm{i}}\textrm{O}_{\textrm{i}}\textrm{H}$)
\cite{Markevich1996} in Si has been obtained from analyses of temperature
dependencies of electron (hole) concentrations.

\begin{figure}
\begin{centering}
\includegraphics[width=8cm]{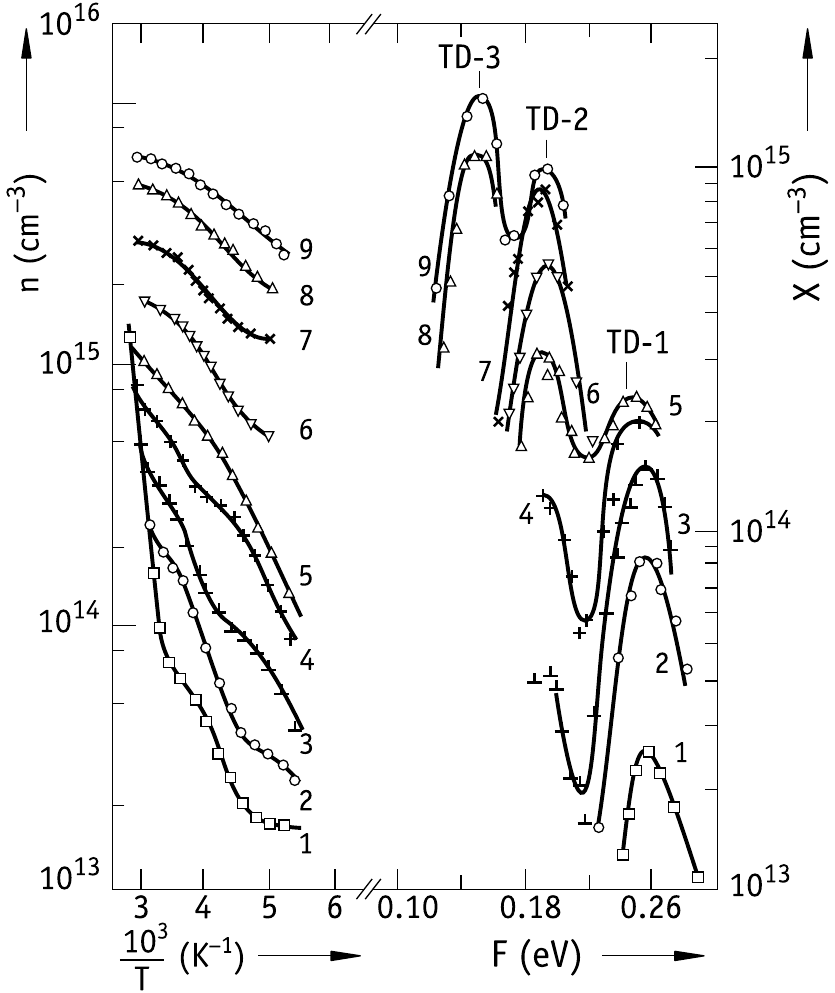}
\par\end{centering}
\caption{\label{FIG:TDD}Temperature dependencies of free electron concentrations,
$n$, (left part) and its derivative $X=k_{\textrm{B}}T(\textrm{d}n/\textrm{d}E_{\textrm{F}})$
(right part) versus the Fermi level with respect to the conduction
band edge $F=E_{\textrm{c}}-E_{\textrm{F}}$ , for an oxygen-rich
Ge crystal subjected to heat-treatments at 623~K for (1) 5~min,
(2) 10~min, (3) 25~min, (4) 60~min, (5) 120~min, (6) 240~min,
(7) 480~min, (8) 1020~min, and (9) 1860~min. (Reproduced with permission
from Litvinov \emph{et~al.} \cite{Litvinov1988}, © 1988, WILEY-VCH
Verlag GmbH \& Co.).}
\end{figure}

The experimental $n\,(p)$ \emph{v.s}. $T$ dependencies can be analysed
either by methods based on solving charge carrier neutrality equations
\cite{Blakemore1987} or by the differential method proposed by Hoffmann
\cite{Hoffmann1979,Hoffmann1982}. According to the differential method,
the concentration of a defect and position of its energy level in
the gap can be determined from a dependency of $X=k_{\textrm{B}}T(\mathrm{d}n/\mathrm{d}E_{\textrm{F}})$
versus $E_{\textrm{F}}$ in semiconductors of n-type {[}$X=k_{\textrm{B}}T(\mathrm{d}p/\mathrm{d}E_{\textrm{F}})$
in semiconductors of p-type{]}. The $X(E_{\textrm{F}})$ dependencies
look like spectra with peaks induced by ionisation of defects. The
magnitude of a peak ($X_{\textrm{m}}$) is proportional to the concentration
of a defect, while the peak position ($E_{\textrm{m}}$) corresponds
to the position of an energy level in the gap. It has been shown in
Refs.~\cite{Hoffmann1979} and \cite{Hoffmann1982} that for defects
with $U>0$, $X_{\textrm{m}}=0.25[\textrm{D}]$ and the half-width
of a band, $\delta E_{\textrm{F}}$, is about $3.5k_{\textrm{B}}T_{\textrm{m}}$,
where $T_{\textrm{m}}$ is the temperature which corresponds to $E_{\textrm{m}}$.
For defects with $U<0$ on the other hand, $X_{\textrm{m}}=[\textrm{D}]$
and $\delta E_{\textrm{F}}\approx1.8k_{\textrm{B}}T_{\textrm{m}}$.
So, an analysis of the half-width of bands in the $X(E_{\textrm{F}})$
dependencies can be considered as a quick test on the sign of effective
correlation energy.

As an example, Fig.~\ref{FIG:TDD} shows $n(10^{3}/T)$ and $X(F)$
dependences with $F=E_{\textrm{c}}-E_{\textrm{F}}$, for an oxygen-rich
Ge crystal, which was subjected to heat-treatments of different durations
at 623~K \cite{Litvinov1988}. Such heat-treatments are known to
result in the introduction of oxygen-related thermal double donors
(TDDs), which consist of a family of subsequently formed (at least
nine) defect species in Ge:O crystals \cite{Clauws2007}. An analysis
of the $X(F)$ spectra in Fig.~\ref{FIG:TDD} showed that the half-width
of the three appearing bands is about $1.8k_{\textrm{B}}T_{\textrm{m}}$,
so giving solid evidence that the first three species of the TDD family
of defects in Ge are centres with negative-$U$ \cite{Litvinov1988}.

From investigations of the occupancy of negative-$U$ defects with
charge carriers at equilibrium conditions, only a value of free energy
for the two-electron charge state change, $\Delta E(N\!\!+\!\!1\,/\,N\!\!-\!\!1)$,
can be determined. Further information about the $E(N\!+\!1\,/\,N)$
and $E(N\,/\,N\!-1\!)$ values, charges states, structural configurations
and energy barriers between them can be obtained from investigations
of non-equilibrium processes of carrier emission and capture.

\subsubsection{Observations of negative-$U$ defects in metastable configurations\label{subsec:obs-negu}}

For the majority of defects with $U<0$, rather large energy barriers
exist between their atomic configurations. Because of these barriers,
under certain conditions, it is possible to `freeze' a negative-$U$
defect in a metastable atomic configuration with a shallower level,
and to obtain information about the electronic properties of the defect
in this configuration. Such `freezing' or, in other words, deep-shallow
excitation can be induced by different methods, \emph{e.g.}, by (i)
injection of minority carriers in certain temperature ranges either
by above-bandgap-energy light pulses or forward bias pulses in n-p
diodes, (ii) a quick change in temperature (quenching), usually from
higher to lower $T$, (iii) cooling down of a semiconductor sample
under external illumination with photons having above-bandgap energies,
(iv) cooling down an n-p diode with an applied reverse bias voltage.
The significant changes in $n\,(p)$ \emph{v.s.} $T$ dependencies,
in DLTS, EPR and various optical (infrared absorption, photoluminescence,
\emph{etc.}) spectra induced by the `freezing'/excitation methods
listed above, usually indicate the presence of metastable (negative-$U$)
defect states in a semiconductor material. From analysis of the appearing
signals in the excited spectra, electronic and structural characteristics
of a negative-$U$ defect in the shallow level configuration can be
determined.

In EPR studies, a number of defects, \emph{e.g.}, positively charged
vacancy and interstitial boron in silicon \cite{Watkins1980,Troxell1980,Harris1982},
Se-related $DX$ centre in AlSb \cite{Stallinga1995} or carbon vacancy
in 4H-SiC \cite{Son2012}, could only be observed in the spectra after
external illumination with above-band-gap light. It has even been
argued in Ref.~\cite{Stallinga1995} that the absence of an EPR signal
in the samples cooled in the dark and its appearance after illumination
can be considered as a direct evidence of the negative-$U$ properties.
It should be noted, however, that such effects can be related to a
defect with a single deep level in the gap and large energy barrier
for capture of a charge carrier.

In Si and Ge crystals containing oxygen-related thermal double donors,
significant changes in $n(T)$ dependencies and IR absorption spectra
have been observed after cooling the investigated samples down under
external illumination with energy $h\nu\geq E_{\textrm{g}}$ \cite{Makarenko1985,Latushko1986,Wruck1986,Litvinov1988,Wagner1989,Murin2003,Clauws2007}.
In the `illuminated' IR absorption spectra of Si crystals, three
sets of absorption bands, which have not been seen in the spectra
recorded after cooling down in the dark, were detected \cite{Latushko1986,Wruck1986,Wagner1989,Murin2003}.
From the analysis of positions of the electronic-transition related
absorption lines in the `illuminated' spectra, the exact locations
of energy levels of three bistable TDD species (negative-$U$ defects)
in the shallow donor configuration could be determined in both Si
and Ge \cite{Latushko1986,Wruck1986,Wagner1989,Murin2003,Clauws2007}.

\begin{figure}
\begin{centering}
\includegraphics[width=7.5cm]{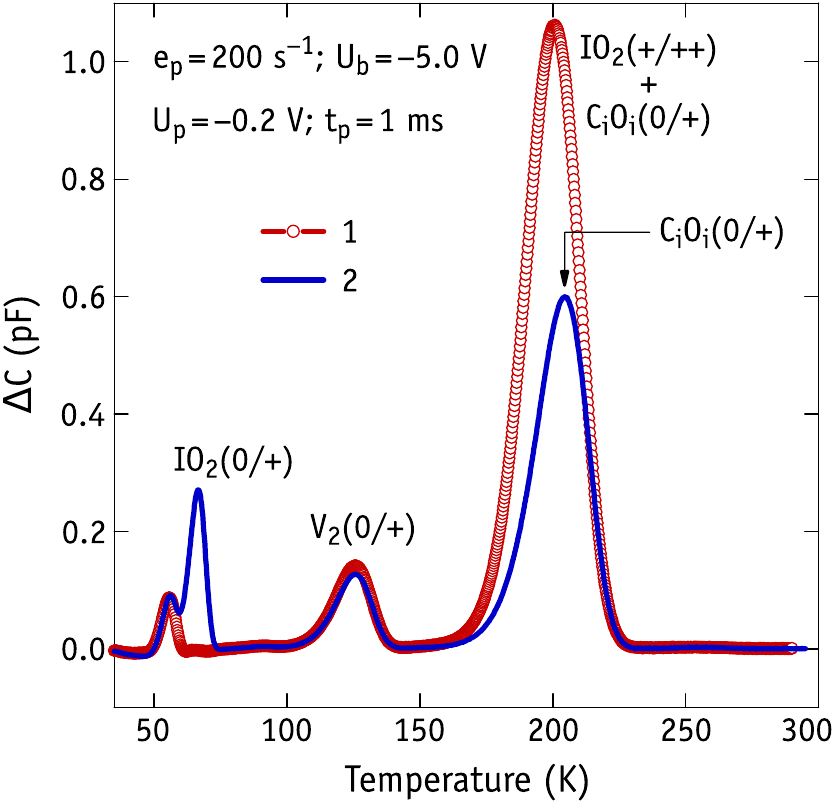}
\par\end{centering}
\caption{\label{FIG:DLTS}DLTS spectra for a boron-doped Cz-Si sample with
low carbon content ($[\textrm{C}]\protect\leq10^{15}\,\textrm{cm}^{-3}$)
after its irradiation with 4~MeV electrons. The dose of irradiation
was $4\times10^{14}\,\textrm{cm}^{-2}$. The spectra were recorded
upon heating the sample from 35~K to 300~K with the measurement
settings shown in the graph. Spectrum 1 was recorded after cooling
the sample down without reverse bias applied to the Schottky diode,
spectrum 2 was recorded after cooling down with the reverse bias on.}
\end{figure}

Changes in DLTS spectra that resulted from different cooling down
conditions of investigated diodes have been reported for semiconductor
samples containing both defects with negative-$U$ properties and
metastable defects with $U>0$ \cite{Chantre1987,Chantre1989,Mukashev2000,Murin2003,Markevich2005a}.
The spectra are usually recorded upon heating the diodes up after
their cooling down either with or without a reverse bias applied.
Cooling down with the applied reverse bias, when there are no free
electrons/holes in the probed `depletion' region, results in `freezing'
of a negative-$U$ defect in a configuration with a shallower level,
which is the minimum energy configuration at high temperatures. So,
typically, in the DLTS spectra recorded after cooling down the diodes
with the applied reverse bias, a signal due to charge emission from
the shallower level appears, and a signal due to emission from the
deeper level either disappears or decreases. As an example, Figure~\ref{FIG:DLTS}
compares the DLTS spectra for a Czochralski silicon (Cz-Si) sample
containing the $I$O$_{2}$ complex (with $I$ standing for a Si self-interstitial).
This complex is a negative-$U$ defect with its first and second donor
levels being at $E(0/+)=E_{\textrm{v}}+0.12$~eV and $E(+/+\!+)=E_{\textrm{v}}+0.36$~eV
\cite{Markevich2005a,Markevich2005b}. The spectra were recorded after
cooling down with either bias-on or bias-off. Cooling down with the
applied reverse bias resulted in the disappearance of the DLTS signal
due to hole emission from the double positively charged state (the
minimum energy configuration at low temperatures) of the $I$O$_{2}$
defect and appearance of the signal due to hole emission from the
metastable single positively charged state.

Further, back transformations from a metastable to stable configuration
can be induced by \emph{e.g.} an increase in temperature if majority
charge carriers are available. From studies of kinetics of these backward
processes at different temperatures in semiconductor crystals with
different free carrier concentrations, a comprehensive set of information
about the electronic structure and concentration of defects with negative-$U$
can be obtained.

\subsubsection{Non-equilibrium occupancy statistics for defects with $U<0$\label{subsubsec:statistics}}

\begin{figure}
\begin{centering}
\includegraphics[width=7.5cm]{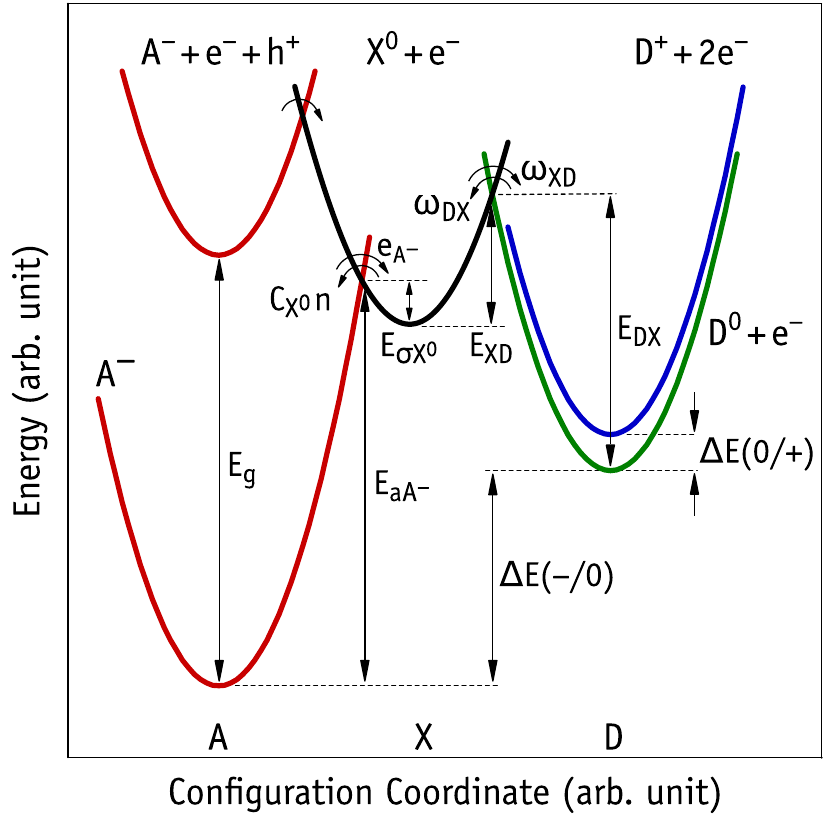}
\par\end{centering}
\caption{\label{FIG:CC-COH}Configuration coordinate diagram for an amphoteric
defect (having an acceptor and a donor levels) with $U<0$.}
\end{figure}

The non-equilibrium occupancy statistics for defects with $U<0$ have
been developed in Refs.~\cite{Tkachev1984,Makarenko1988,Markevich1997}.
For an elucidation of the details of these statistics let us consider
a total concentration $N_{\textrm{d}}$ of an amphoteric defect (having
acceptor and donor levels) with negative-$U$ level ordering, which
exchange charge carriers (electrons) with the conduction band. Figure~\ref{FIG:CC-COH}
shows a general configuration-coordinate diagram for such a defect.
In the absence of minority carriers, transitions between the stable
acceptor ($\textrm{A}^{-}$) and donor ($\textrm{D}^{+}$) states
occur through the metastable $\textrm{X}^{0}$ and $\textrm{D}^{0}$
states according to the following sequence of reactions:

\begin{equation}
\textrm{A}^{-}\rightleftarrows\textrm{A}^{0}+\textrm{e}^{-}\rightleftarrows\textrm{X}^{0}+\textrm{e}^{-}\rightleftarrows\textrm{D}^{+}+2\textrm{e}^{-}\label{EQ:EXP4}
\end{equation}
The changes in the density of defect states and in the free electron
concentration can be described by the following set of differential
equations,

\begin{eqnarray*}
\frac{\textrm{d}[\textrm{A}^{-}]}{\textrm{d}t} & = & +c_{\textrm{X}^{0}}[\textrm{X}^{0}]-e_{\textrm{A}^{-}}[\textrm{A}^{-}],\\
\frac{\textrm{d}[\textrm{X}^{0}]}{\textrm{d}t} & = & -c_{\textrm{X}^{0}}[\textrm{X}^{0}]+e_{\textrm{A}^{-}}[\textrm{A}^{-}]-\omega_{\textrm{XD}}[\textrm{X}^{0}]+\omega_{\textrm{DX}}[\textrm{D}^{0}],\\
\frac{\textrm{d}[\textrm{D}^{0}]}{\textrm{d}t} & = & +c_{\textrm{D}^{+}}[\textrm{D}^{+}]+e_{\textrm{D}^{0}}[\textrm{D}^{0}]+\omega_{\textrm{XD}}[\textrm{X}^{0}]-\omega_{\textrm{DX}}[\textrm{D}^{0}],\\
\frac{\textrm{d}[\textrm{D}^{+}]}{\textrm{d}t} & = & -c_{\textrm{D}^{+}}[\textrm{D}^{+}]+e_{\textrm{D}^{0}}[\textrm{D}^{0}],\\
\frac{\textrm{d}n}{\textrm{d}t} & = & -c_{\textrm{X}^{0}}[\textrm{X}^{0}]+e_{\textrm{A}^{-}}[\textrm{A}^{-}]-c_{\textrm{D}^{+}}[\textrm{D}^{+}]+e_{\textrm{D}^{0}}[\textrm{D}^{0}],
\end{eqnarray*}
where emission and capture rates, $e_{Q}$ and $c_{Q}$, are defined
as

\begin{equation}
e_{Q}=c_{Q'}(N_{\textrm{c}}/n)\exp[-\Delta E(Q/Q')/k_{\textrm{B}}T]
\end{equation}
and

\begin{equation}
c_{Q}=\sigma_{\infty Q}n\langle v_{\textrm{c}}\rangle_{\textrm{th}}\exp(-E_{\sigma Q}/k_{\textrm{B}}T),
\end{equation}
respectively, with $Q$ and $Q'$ referring to neighbouring configurations
with different charge. Structural transformations $\textrm{X}\rightleftarrows\textrm{D}$
without a charge state change, occur at a rate $\omega_{QQ'}=\omega_{\infty,QQ'}\exp(-E_{QQ'}/k_{\textrm{B}}T)$,
where now $Q$ and $Q'$ refer to neutral X or D states. Defect concentrations
$[Q^{q}]$ of configuration $Q$ in charge state $q$, are subject
to $\sum_{q}[Q^{q}]=N_{\textrm{d}}$. For the transformation process,
$\omega_{\infty,QQ'}$ and $E_{QQ'}$ are the high-temperature attempt
frequency and transformation barrier, respectively.

The full solution of the above set of the differential equations is
not an easy task. It should be noted, however, that some of the reaction
rates are usually much faster than others, so some of the states are
in quasi-equilibrium. Furthermore, upon the assumption that $N_{\textrm{d}}\ll n\approx$~`const',
the above set of equations can be transformed to a set of first order
differential equations, which can be solved analytically. Below we
will consider in more detail the changes in the density of the $\textrm{A}^{-}$
state, which is usually monitored in DLTS measurements. The changes
in the density of the $\textrm{A}^{-}$ state at a constant temperature
after an initial deviation from equilibrium, $\Delta[\textrm{A}^{-}]_{0}$,
can be expressed by the following equation,

\begin{equation}
\Delta[\textrm{A}^{-}](t)=\Delta[\textrm{A}^{-}]_{0}\exp(-t/\tau),\label{EQ:EXP5}
\end{equation}
where $\tau$ is the characteristic time of the decay process. The
characteristic time depends on state transition rates as \cite{Markevich1997}:

\begin{equation}
\tau^{-1}=\frac{\omega_{\textrm{XD}}e_{\textrm{A}^{-}}+c_{\textrm{X}^{0}}\omega_{\textrm{DX}}f_{\textrm{D}}}{\omega_{\textrm{DX}}+c_{\textrm{X}^{0}}},\label{EQ:EXP6}
\end{equation}
where

\begin{eqnarray}
f_{\textrm{D}} & = & \frac{[\textrm{D}^{0}]}{[\textrm{D}^{0}]+[\textrm{D}^{+}]}=\left[1+\exp\!\left(-\frac{E_{\textrm{F}}\!-\!E(0/+)}{k_{\textrm{B}}T}\right)\right]^{-1},\label{EQ:EXP6-1}\\
 & = & \left[1+\frac{N_{\textrm{c}}}{n}\exp\left(-\frac{\Delta E(0/+)}{k_{\textrm{B}}T}\right)\right]^{-1},\label{EQ:EXP6-2}
\end{eqnarray}
is the occupancy function for the defect in the donor configuration.
Equation~\ref{EQ:EXP6} can be expressed in a common way as

\begin{equation}
\tau^{-1}=e_{\textrm{eff}}+c_{\textrm{eff}},\label{EQ:EXP7}
\end{equation}
with

\begin{equation}
e_{\textrm{eff}}=\frac{e_{\textrm{A}^{-}}}{1+c_{\textrm{X}^{0}}/\omega_{\textrm{XD}}},\label{EQ:EXP8}
\end{equation}
and

\begin{equation}
c_{\textrm{eff}}=\frac{c_{\textrm{X}^{0}}\omega_{\textrm{DX}}\omega_{\textrm{XD}}^{-1}f_{\textrm{D}}}{1+c_{\textrm{X}^{0}}/\omega_{\textrm{XD}}}.\label{EQ:EXP9}
\end{equation}

An analysis of Eqs.~\ref{EQ:EXP6}, \ref{EQ:EXP7}, \ref{EQ:EXP8}
and \ref{EQ:EXP9} indicates that several terms, with their specific
activation energies and power dependence on the free carrier concentration,
can dominate the temperature dependence of $\tau^{-1}$. Particularly
influential factors are the Fermi level position with respect to $E(-/0)$,
$E(0/+)$ and $E(-/+)$, and the $c_{\textrm{X}^{0}}/\omega_{\textrm{XD}}$
ratio. The effective emission rate, $e_{\textrm{eff}}$, is the dominant
term in Eq.~\ref{EQ:EXP7} when the Fermi level is below the $E(-/+)=E_{\textrm{c}}-[E(-/0)+E(0/+)]/2$
occupancy level of a negative-$U$ defect. At these conditions, $\tau^{-1}=e_{\textrm{eff}}$.
When $c_{\textrm{X}^{0}}\ll\omega_{\textrm{XD}}$, $\tau^{-1}=e_{\textrm{A}^{-}}$,
so the temperature dependence of $\tau^{-1}$ can be described by
a commonly used equation for single electron emission. For the cases
of $c_{\textrm{X}^{0}}\gg\omega_{\textrm{XD}}$, we have $\tau^{-1}=\omega_{\textrm{XD}}e_{\textrm{A}^{-}}/c_{\textrm{X}^{0}}$,
so the transformation rate (actually the rate of occupancy of the
$\textrm{A}^{-}$ state) is inversely proportional to the free electron
concentration. Such unusual dependence of $\tau^{-1}$ versus $n$
has been clearly observed for a complex consisting of a substitutional
boron atom and oxygen dimer in silicon \cite{VaqueiroContreras2019,Markevich2019}.

The capture term in Eq.~\ref{EQ:EXP7} is dominant when $E_{\textrm{F}}>E(-/+)$.
In this case, $\tau^{-1}=c_{\textrm{eff}}$ and up to four specific
terms with different values of activation energy and power dependence
on $n$ can occur, depending on the $c_{\textrm{X}^{0}}/\omega_{\textrm{XD}}$
ratio and the Fermi level position with respect to $E(0/+)$. For
a number of defects, \emph{e.g.} bistable TDDs in Si and Ge crystals,
under certain conditions, $\tau^{-1}\sim n^{2}$ dependencies have
been observed \cite{Tkachev1984,Litvinov1988,Makarenko1988,Markevich2018}.
Such dependencies clearly show a two-fold change in the charge state
of a defect upon capture of carriers, so indicating its negative-$U$
properties.

From the analysis of temperature dependencies of transition rates
between configurations of a defect with $U<0$ monitored by measurements
of changes in either free carrier concentration or density of a specific
defect state in semiconductor crystals with different $n\,(p)$, it
is possible to elucidate the electronic structure of the defect and
determine practically all its energy differences and transformation
barriers. This has been done for a number of defects in Ge and Si
crystals \cite{Tkachev1984,Litvinov1988,Makarenko1988,Markevich1997,Markevich2005a,Markevich2018,VaqueiroContreras2019,Markevich2019}.

\subsection{Modelling negative-$U$ defects\label{subsec:modelling}}

\subsubsection{Calculation of electronic transition levels\label{subsubsec:levels}}

The calculation of electronic transition levels is essentially a problem
of finding the energy of an electronic reservoir (Fermi level), for
which the exchange of electrons with a defective sample becomes energetically
favourable (see Section~\ref{subsec:levels}). This is a topic which
has been widely revised in the past (see for instance Ref.~\cite{Todorova2014}
and \cite{Drabold2007}), so we leave here the essential features.
We start by recalling Eq.~\ref{EQ:EFRM2}, which determines the energy
needed to create a defect in a crystal (defect formation energy)

\begin{equation}
E_{\textrm{f}}^{(q)}(\mathbf{R},\mu_{i},E_{\textrm{F}})=E^{(q)}(\mathbf{R})-\sum_{i}n_{i}\mu_{i}+qE_{\textrm{F}}.\label{EQ:EFRM3}
\end{equation}

Here, $E^{(q)}(\mathbf{R})$ is the energy of a large portion of a
sample enclosing a single defect, including its electronic wave functions.
The defective sample volume is made of $n_{i}$ atoms of species $i$
with chemical potential $\mu_{i}$ and collective atomic coordinate
$\mathbf{R}=\{\mathbf{R}_{\alpha}\}$ ($\alpha$ being an atomic index).
It can be easily shown that a transition energy level between charges
states $q$ and $q'$ (with $q$ being more negative) is located at

\begin{equation}
E(q/q')=-\frac{E^{(q)}(\mathbf{R})-E^{(q')}(\mathbf{R}')}{q-q'}.\label{EQ:LEVEL2}
\end{equation}
Equation~\ref{EQ:LEVEL2} accounts for the fact that charge states
$q$ and $q'$ may correspond to radically different atomistic geometries
$\mathbf{R}$ and $\mathbf{R}'$.

The large chunk of material with a defect referred above is normally
approximated to a defective supercell. The inherent periodic boundary
conditions imply the cell to be neutral, even when we add (remove)
electrons to (from) a defect state located in the gap. Technically,
a uniform background counter-charge of density $-q/\Omega$ (with
$\Omega$ being the supercell volume) is superimposed to the electronic
density. This implies that the aperiodic energy $E^{(q)}$ that enters
in Equations~\ref{EQ:EFRM3} and \ref{EQ:LEVEL2}, is related to
the periodic energy $\widetilde{E}^{(q)}$ obtained from the supercell
calculation as $E^{(q)}=\widetilde{E}^{(q)}+E_{\textrm{pcc}}^{(q)}$,
where $E_{\textrm{pcc}}^{(q)}$ is a correction often referred to
as \emph{periodic charge correction}. Several schemes have been proposed
for the calculation of $E_{\textrm{pcc}}^{(q)}$ (see for instance
Ref.~\cite{Komsa2012} and references therein).

Of course, in a semiconductor, transitions are only observable if
they are located in the range $E_{\textrm{v}}<E(q/q')<E_{\textrm{c}}$.
We have therefore to calculate $E_{\textrm{v}}$ or $E_{\textrm{c}}$
in order to cast the levels as measurable quantities, $i.e.$, $E(q/q')-E_{\textrm{v}}$
or $E_{\textrm{c}}-E(q/q')$. One possibility is to assume that $E_{\textrm{\{v,c\}}}=\epsilon_{\{\textrm{v,c}\},\textrm{bulk}}$,
which are the highest occupied ($\epsilon_{\textrm{v},\textrm{bulk}}$)
and lowest unoccupied ($\epsilon_{\textrm{c},\textrm{bulk}}$) states
of a single-particle Hamiltonian (like the Kohn-Sham equations). Another
option is to follow the $\Delta$SCF (delta self-consistent field)
method to obtain $E_{\textrm{v}}=\widetilde{E}_{\textrm{bulk}}^{(0)}-\widetilde{E}_{\textrm{bulk}}^{(+)}$
or $E_{\textrm{c}}=\widetilde{E}_{\textrm{bulk}}^{(-)}-\widetilde{E}_{\textrm{bulk}}^{(0)}$
from total energies of bulk supercells. These must be identical in
size and shape to those used to obtain $\widetilde{E}^{(q)}(\mathbf{R})$
values. In any case, we should be aware that the accuracy of the calculated
defect levels, including the energies of the band gap edges, strongly
depends on the quality of the Hamiltonian. Particularly important
is the level of detail put in the description of the exchange and
correlation interactions between electrons (see for instance Ref.~\cite{Alkauskas2008}).

\subsubsection{Calculation of carrier capture cross-sections in multi-phonon assisted
transitions\label{subsubsec:capture}}

Inelastic scattering of free carriers by defects essentially consists
of the transfer of energy and momentum of a propagating electron or
hole, to another carrier bound to a defect (trap-Auger process), its
conversion into radiation (luminescence process), into nuclear motion
(\emph{e.g.} capture enhanced defect migration) or heat dissipation
(multi-phonon emission process) \cite{Stoneham2001}. Non-radiative
capture of carriers via multi-phonon emission (MPE) is therefore a
special case of inelastic scattering -- the energy drop of the traveling
carrier is fully dissipated into atomic vibrations. This is the most
common capture mechanism at deep traps involving negative-$U$ defects.
Other important processes, like recombination, generation or emission,
can be described either as a sequence of consecutive capture events,
or by reversing the operation using detailed balance.

The relevant quantity to be evaluated is the carrier capture rate,
$c_{n/p}$, which relates to the apparent capture cross-section of
Eq.~\ref{EQ:CCSEC} as $c_{n}=\sigma_{n}\langle v_{\textrm{c}}\rangle{}_{\textrm{th}}\,n$
for the case of electrons (with an obvious analogous expression for
holes). While first-principles calculations of electronic transition
levels have been routinely reported in the literature, the calculation
of capture rates is clearly lagging behind. For a historical account
regarding the development MPE theory, we divert the reader to Refs.~\cite{Stoneham2001,Alkauskas2014,Barmparis2015}.
In general, the calculation of the non-radiative transition rate between
free and bound states starts with the description of a many-body Hamiltonian
$\hat{H}$ and the choice of a practical basis to describe the total
initial and final wave functions involved, $\Psi_{\textrm{i}}$ and
$\Psi_{\textrm{f}}$, respectively. The initial state represents the
defect plus uncorrelated free carrier, whereas the final state stands
for the trapped state. The \emph{static approximation} is a rather
convenient starting point in the context of popular electronic structure
methods like Hartree-Fock and density functional theory. Here, the
electronic wave functions $\psi_{s;\mathbf{R}_{0}}(\mathbf{r})$ ($s$
being an electronic state index and $\mathbf{r}$ all electronic degrees
of freedom) are found from a static Hamiltonian $\hat{H}_{\mathbf{R}_{0}}$,
constructed for a fixed atomic geometry $\mathbf{R}_{0}$. The nuclear
wave functions are treated separately. Assuming the harmonic approximation,
we write the initial state as $\Psi_{\textrm{i}m}=\psi_{\textrm{i};\mathbf{R}_{0}}(\mathbf{r},\{Q_{\textrm{i}k}\})\chi_{\textrm{i}m}(\{Q_{\textrm{i}k}\})$
and likewise, the final state as $\Psi_{\textrm{f}m}=\psi_{\textrm{f};\mathbf{R}_{0}}(\mathbf{r},\{Q_{\textrm{f}k}\})\chi_{\textrm{f}n}(\{Q_{\textrm{f}k}\})$.
Here, $m$ and $n$ are quantum numbers for the vibrational states
$\chi_{\textrm{i}}$ and $\chi_{\textrm{f}}$, respectively, while
$k$ indexes are used to identify the $3N$ individual vibrational
modes $Q_{\{\textrm{i,f}\}k}$, $N$ being the number of atoms in
the sample.

The transition probability between states $\Psi_{\textrm{i}m}$ and
$\Psi_{\textrm{f}n}$ relates to the off-diagonal matrix elements
of the Hamiltonian when perturbed by a change in geometry (away from
$\mathbf{R}_{0}$) induced by the \emph{promoting modes} $\{Q_{\textrm{i}k}\}$
of the initial electronic state. Such quantities may be obtained by
first-order expansion of $\hat{H}$ in a Taylor series of $Q_{\textrm{i}k}$
(\emph{linear coupling approximation}),

\begin{equation}
\Delta H_{\textrm{i}m,\textrm{f}n}=\sum_{k}\left\langle \psi_{\textrm{i},\mathbf{R}_{0}}\right|\frac{\partial\hat{H}}{\partial Q_{\textrm{i}k}}\left|\psi_{\textrm{f},\mathbf{R}_{0}}\right\rangle \left\langle \chi_{\textrm{i}m}\right|\Delta Q_{\textrm{i}k}\left|\chi_{\textrm{f}m}\right\rangle ,\label{EQ:MATEL}
\end{equation}
where $\Delta Q_{\textrm{i}k}=Q_{\textrm{i}k}-Q_{0k}$ with $Q_{0k}$
representing a generalised coordinate, equivalent to the collective
nuclear coordinate $\mathbf{R}_{0}$ in the $\{Q_{\textrm{i}k}\}$
basis. The transition (capture) rate is then cast in the form of Fermi’s
golden rule,

\begin{equation}
c=\frac{2\pi}{\hbar}\sum_{m,n}w_{m}(T)|\Delta H_{\textrm{i}m,\textrm{f}n}|^{2}\delta(\Delta E_{\textrm{i}m,\textrm{f}n}),\label{EQ:FGR1}
\end{equation}
which sums up all possible vibronic transitions, weighted by the occupancy
fraction $w_{m}$ of the promoting vibronic states. Under thermal
equilibrium $w_{m}$ is simply a normalised Boltzmann distribution.
The allowed transitions in Equation~\ref{EQ:FGR1} are also restricted
to those which conserve the energy, hence the use of a Dirac delta
$\delta(\Delta E_{\textrm{i}m,\textrm{f}n})=\delta(E_{\textrm{i}}-E_{\textrm{f}}+m\hbar\omega_{m}-n\hbar\omega_{n})$.
In practice, the $\delta$ function is replaced by a normalised Gaussian
function with full width $\sim k_{\textrm{B}}T$ to account for thermal
broadening.

Up to this point, several bold approximations were already made. Still,
the formulation of $\Delta H_{\textrm{i}m,\textrm{f}n}$ as its stands
in Eq.~\ref{EQ:MATEL}, makes its calculation a rather cumbersome
exercise. The problem lies essentially on the identification of the
vibrational modes which lead to non-vanishing $\langle\chi_{\textrm{i}}|\Delta Q_{\textrm{i}k}|\chi_{\textrm{f}}\rangle$
terms. To give an idea, using a triply hydrogenated vacancy ($V$H$_{3}$)
defect in silicon as a benchmark, Barmparis and co-workers \cite{Barmparis2015}
used a Monte Carlo sampling scheme, being able to find convergence
for $\Delta H_{\textrm{i}m,\textrm{f}n}$ when about 12 distinct phonon
modes, among millions, were considered.

Motivated by the fact that a judicious choice of a single effective
phonon mode resulted in a promising agreement between the observed
and calculated luminescence intensity of radiative transitions (see
for instance Refs.~\cite{Schanovsky2011} and \cite{Alkauskas2012}),
several authors evaluated the transition matrix elements within the
single-phonon approximation. Accordingly, the effective mode is the
one that maximises the coupling to the distortion during the capture
process,

\begin{equation}
Q^{2}=\sum_{\alpha}m_{\alpha}\lambda^{2}|\Delta\mathbf{R}_{\alpha}|^{2}.\label{EQ:MODE}
\end{equation}

Equation~\ref{EQ:MODE} is a linear interpolation between initial
and final atomic coordinates $\Delta\mathbf{R}_{\alpha}=\mathbf{R}_{\textrm{f},\alpha}-\mathbf{R}_{\textrm{i},\alpha}$
with $\alpha$ running over all $N$ atoms with mass $m_{\alpha}$.
The displacement amplitude along the effective mode is governed by
a unitless factor $0\leq\lambda\leq1$. The units of $Q$ are amu$^{1/2}$~Å
(amu being the atomic mass unit). From the effective mode we arrive
at the corresponding effective vibrational frequency $\omega_{\{\textrm{i,f}\}}$
of the initial or final states by extracting $\omega_{\{\textrm{i,f}\}}=\partial E_{\{\textrm{i,f}\}}/\partial Q$
from total energy calculations. With this in mind, Equation~\ref{EQ:FGR1}
simplifies into a factorisation of purely electronic and vibrational
terms,

\begin{equation}
c=\frac{2\pi}{\hbar}W_{\textrm{if}}^{2}\,G_{\textrm{if}}(T),\label{EQ:FGR2}
\end{equation}
where the transition matrix element is now $W_{\textrm{if}}=\langle\psi_{\textrm{i}}|\partial\hat{H}/\partial Q|\psi_{\textrm{f}}\rangle$
and $G_{\textrm{if}}$ is the temperature-dependent line shape factor

\begin{equation}
G_{\textrm{if}}(T)=\sum_{m,n}w_{m}(T)|\langle\chi_{\textrm{i}m}|\chi_{\textrm{f}n}\rangle|^{2}\delta(\Delta E_{\textrm{i}m,\textrm{f}n}).\label{EQ:LINSHP}
\end{equation}

Equations~\ref{EQ:FGR2} and \ref{EQ:LINSHP} translate the \emph{Franck-Condon
approximation}, which separates an instantaneous electronic transition
from the \emph{sluggish} phonon dissipation process that follows.
The calculation of $G_{\textrm{if}}$ boils down to the Frank-Condon
integrals $\langle\chi_{\textrm{i}m}|\chi_{\textrm{f}n}\rangle$ for
all vibrational promoting and accepting states. For that, recursive
techniques have been proposed \cite{Manneback1951,Doktorov1977,Zapol1982,Schmidt2010}
and applied on different contexts \cite{Borrelli2003,Schanovsky2011,Krasikov2016}.
Sampling methods have also been applied \cite{Barmparis2015}, although
inclusion of a few phonons implied the evaluation of millions of $\langle\chi_{\textrm{i}m}|\chi_{\textrm{f}n}\rangle$
configurations that matched the electronic energy off-set.

Regarding the evaluation of $W_{\textrm{if}}$, different approaches
have also been used. Alkauskas \emph{et~al.} \cite{Alkauskas2014}
replaced the many-body Hamiltonian and wave functions by single-particle
counterparts $\hat{h}$ and $\phi_{\{\textrm{i,f}\}}$, which were
taken from the Kohn-Sham equations of density-functional theory. Thus,
$W_{\textrm{if}}$ becomes

\begin{equation}
W_{\textrm{if}}=\langle\phi_{\textrm{i}}|\partial\hat{h}/\partial Q|\phi_{\textrm{f}}\rangle,\label{EQ:WKS}
\end{equation}
which can be solved by first-order perturbation theory \cite{Alkauskas2014}.
Alternatively, a phenomenological approach was pursued by Krasikov
\emph{et~al.} \cite{Krasikov2016} who used Eq.~88 in the paper
of Henry and Lang \cite{Henry1977},

\begin{equation}
|W_{\textrm{if}}|^{2}=\frac{2\pi}{\Omega}\left(\frac{\hbar^{2}}{2m^{*}}\right)^{3/2}\epsilon^{1/2},\label{EQ:WEL}
\end{equation}
where $\Omega$ is the crystal volume, $m^{*}$ is the effective mass
of the free carrier and $\epsilon$ is the energy separation between
the initial and final states near the intersection point (see discussion
in Section~\ref{subsec:ccdiagram}).

\subsubsection{Calculation of transformation barriers\label{subsubsec:barriers}}

Defect reorientation and transformation are important processes that
occur during the experimental characterisation of negative-$U$ defects.
Understanding the underlying physics of such processes is fundamental
in order to explain and ultimately to control defect behaviour. The
problem is usually addressed in the spirit of transition state theory
\cite{Flynn1972}, and usually involves finding the lowest free energy
barrier that separates specific reactants from products, or in the
case of defects, initial from final structures.

Although being a rather challenging problem, particularly when we
have no clue about the mechanisms involved, the search for a saddle-point
along a potential energy landscape can be investigated in a number
of ways. In general, defect transformation mechanisms are investigated
via constrained-relaxation of the atomistic structure, ideally using
a highly accurate quantum-mechanical method to evaluate the forces.
Among the existing algorithms we highlight:

The \emph{dimer method} \cite{Henkelman1999}, which requires knowledge
of a single minimum energy configuration, performs a search for a
nearby saddle point. This method is particularly useful to look for
saddle points along paths with unknown final configurations. The dimer
method works on two structures (the \emph{dimer}) separated by a small
distance in configurational space. The direction along the lowest
curvature of the potential energy is obtained by rotation of the dimer,
and by finding the forces for each structure. This result is used
to move the dimer uphill, from a stable configuration to a near saddle
point.

The \emph{Lanczos method} was originally referred to as ART algorithm
by Malek and Mousseau \cite{Malek2000} and is a close relative to
the dimer method. It can be used to search for a saddle point close
to an arbitrary initial configuration. Unlike the dimer method, the
Lanczos scheme estimates the Hessian eigenvector that corresponds
to the lowest eigenvalue by expansion of the potential energy surface
in a Krylov subspace. After this stage, the method proceeds in the
same way as the dimer.

The \emph{nudged elastic band} (NEB) \emph{method} \cite{Henkelman2000a,Henkelman2000b}
is a robust algorithm that works when the initial and final states
of the mechanism are known \emph{a priori}. It basically involves
the relaxation of a structure sequence (referred to as intermediate
images) along the multi-dimensional path that separates the end-structures.
Usually a starting guess for the intermediate images is obtained by
linear interpolation of the end-structures. The latter are kept fixed
during the relaxation process. During the relaxation of each image,
equal spacing between neighbouring images is maintained by inclusion
of artificial spring forces.

\begin{figure}
\begin{centering}
\includegraphics[width=8cm]{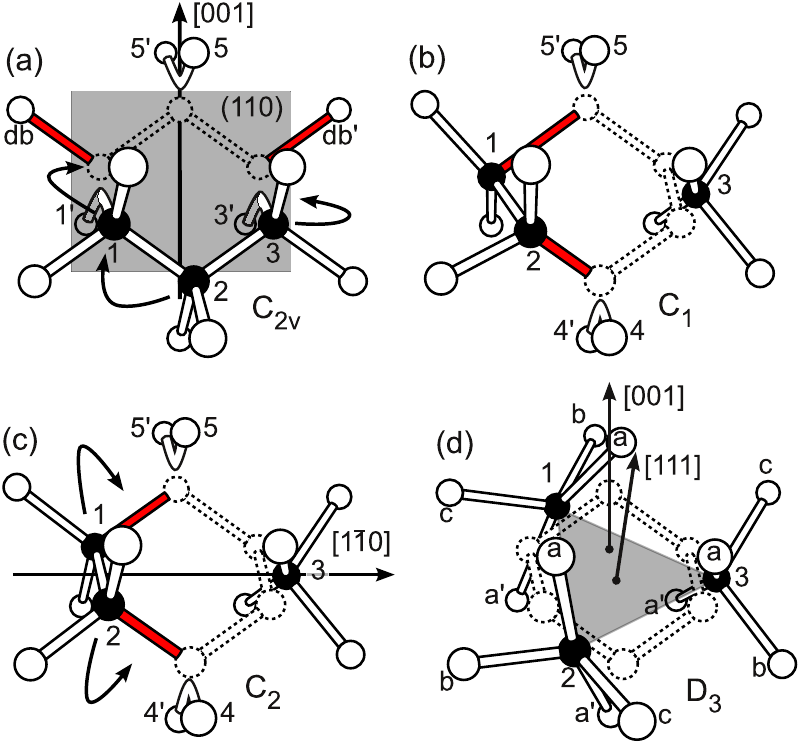}
\par\end{centering}
\caption{\label{FIG:V3-REORIENT}Atomic structure of stable and metastable
trivacancy defects in silicon. The transformation mechanism is depicted
in four stages (a), (b), (c) and (d) by the arrows indicating the
motion of three Si atoms at the core (shown in black for a better
perception). Perfect lattice sites are drawn with a dashed line. Broken
bonds are shown as solid red sticks. (Reprinted with permission from
Ref.~\cite{Coutinho2012}. © 2012 by the American Physical Society).}
\end{figure}

An example of a defect-related mechanism studied using the NEB method
is depicted in Figure~\ref{FIG:V3-REORIENT}, showing the transformation
of the bistable trivacancy ($V_{3}$) complex in Si \cite{Coutinho2012}.
This is one of the most important radiation products in heavy particle
irradiated Si \cite{Markevich2009}. Right after irradiation the observed
defect has the structure depicted in Figure~\ref{FIG:V3-REORIENT}(a).
The Si broken bonds on this structure are responsible for several
deep donor and acceptor levels observed by DLTS \cite{Markevich2009}.
However, storage for a few weeks of the irradiated Si samples at room
temperature result in the transformation to the structure of Figure~\ref{FIG:V3-REORIENT}(d).
This is accompanied by the disappearance of the deep levels and the
introduction of shallow electron traps at 75~meV bellow the conduction
band bottom.

The energy barrier for the above transformation was estimated using
the NEB method as 1.15~eV and 1.14~eV for neutral and negatively
charged $V_{3}$, respectively \cite{Coutinho2012}. These figures
match very well their respective experimental counterparts of 1.16~eV
and 1.15~eV \cite{Markevich2009}.

\section{Showcase of negative-$U$ defects and their characterisation\label{sec:showcase}}

In this section we provide some examples of negative-$U$ centres
in widely used semiconductors. In the selection we try to illustrate
a range of behaviour, the convergence of theory and experiment, and
cases which have provided particular challenges in understanding.
Among this selection are examples of negative-$U$ centres which are
of importance in semiconductor devices. The section is divided in
two parts, firstly mini reviews of published work on four defects,
1) $\mathrm{B_{s}O_{2}}$ (LID) complex in silicon, 2) atomic hydrogen
in silicon, 3) $DX$ centres in III-V alloys and 4) The Ga-$\textrm{O}_{\textrm{As}}$-Ga
defect in GaAs. This is followed by a table which covers a wider range
of important negative-$U$ centres providing basic parameters and
key references. The table includes the four defects listed above and
negative-$U$ defects referred to previously in the text to illustrate
particular aspects of negative-$U$ behaviour, and theoretical and
experimental challenges in their study.

\subsection{$B_{s}O_{2}$ (LID) complex in silicon\label{subsec:bo2-lid}}

A complex consisting of a substitutional boron atom and an oxygen
dimer ($\mathrm{B_{s}O_{2}}$) in Si was suggested to be responsible
for light induced degradation (LID) of solar cells produced from silicon
crystals, which contain boron and oxygen impurity atoms \cite{Schmidt2004,Bothe2006,Niewelt2017}.
The suggestion was based on the experimental observations of linear
and quadratic dependencies of concentration of recombination active
defects responsible for LID on substitutional boron and interstitial
oxygen concentrations, respectively \cite{Schmidt2004,Bothe2006}.
However, for almost two decades, there was no clear understanding
of the electronic structure of the complex or the mechanisms of its
formation and transformations \cite{Niewelt2017}. Answers to the
above uncertainties have been found recently in a complex study consisting
of a range of experimental techniques (various junction spectroscopy
methods, photoluminescence, microwave detected photo-conductance decays)
and \emph{ab-initio} modelling \cite{VaqueiroContreras2019,Markevich2019},
where it was concluded that $\mathrm{B_{s}O_{2}}$ has negative-$U$
properties.

\begin{figure}
\begin{centering}
\includegraphics[width=8cm]{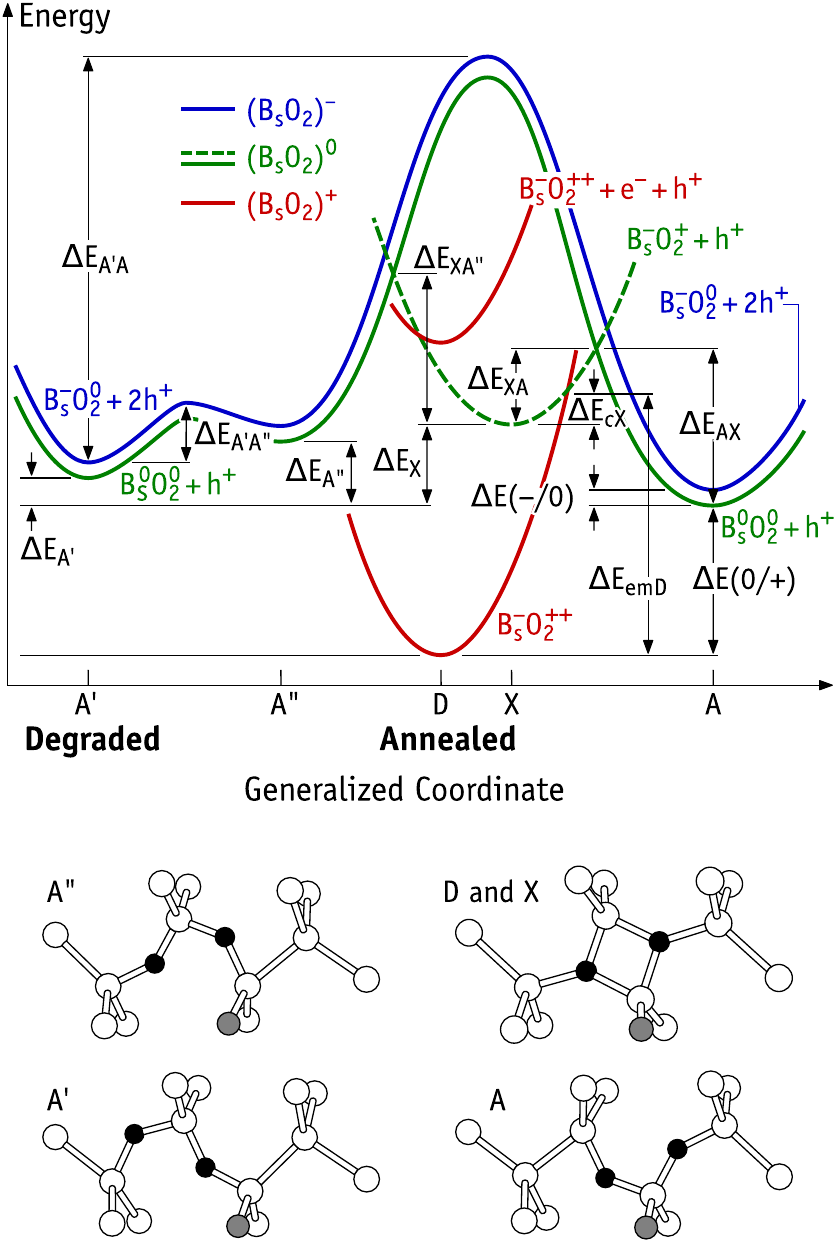}
\par\end{centering}
\caption{\label{FIG:BO2}Configuration coordinate diagram and atomic configurations
of the $\textrm{B}_{\textrm{s}}\textrm{O}_{2}$ complex in Si on different
charge states. (Reprinted from Ref.~\cite{VaqueiroContreras2019},
with the permission of AIP Publishing).}
\end{figure}

Figure~\ref{FIG:BO2} shows the configuration-coordinate diagram
and atomic configurations of the $\mathrm{B_{s}O_{2}}$ complex in
different charge states. The negative-$U$ properties of the complex
are related to the transformations of the oxygen dimer in the vicinity
of a substitutional boron atom between the so called `squared' configuration
(D and X in Figure~\ref{FIG:BO2}, with two three-fold coordinated
oxygen atoms) and `staggered' configurations (configurations $\textrm{A}$,
$\textrm{A}'$, and $\textrm{A}''$ in Figure~\ref{FIG:BO2}, with
two-fold coordinated oxygen atoms). In p-type Si crystals when $E_{\textrm{F}}<E_{\textrm{v}}+0.3$~eV,
the $\mathrm{B_{s}O_{2}}$ centre is in the positive charge state
(state $\textrm{D}^{+}$ in Figure~\ref{FIG:BO2}). In this state,
the defect can either capture an electron (created by either light
or forward bias pulses) or emit a hole (thermally when the temperature
is high enough) and transform to the shallow acceptor $\textrm{A}^{-}$
state according to the following sequence of transitions: $\textrm{D}^{+}+\textrm{h}^{+}+\textrm{e}^{-}\rightleftarrows\textrm{X}^{0}+\textrm{h}^{+}\rightleftarrows\textrm{A}^{0}+\textrm{h}^{+}\rightleftarrows\textrm{A}^{-}+2\textrm{h}^{+}$.
Kinetics of the forward and back transitions between the $\textrm{D}^{+}$
and $\textrm{A}^{-}$ have been monitored at different temperatures
in the p-type Si crystals with different hole concentrations and analysed
with the use of Equations~\ref{EQ:EXP7}, \ref{EQ:EXP8} and \ref{EQ:EXP9}
\cite{VaqueiroContreras2019,Markevich2019}. From this analysis, the
values of energy differences and barriers between different configurations
of the complex have been derived.

Further, it has been found that upon relatively long (tens of hours
at room temperature) minority-carrier-injection treatments the $\mathrm{B_{s}O_{2}}$
defect transforms to a metastable configuration with a shallow acceptor
level and stronger recombination activity (than that for $\mathrm{B_{s}}$
and the A state of the $\mathrm{B_{s}O_{2}}$ complex). A possible
mechanism of this transformation has been considered in Ref.~\cite{VaqueiroContreras2019}.
\emph{Ab-initio} modelling results have shown the existence of two
metastable shallow acceptor configurations ($\textrm{A}'$ and $\textrm{A}''$
configurations in Fig.~\ref{FIG:BO2}) of the $\mathrm{B_{s}O_{2}}$
complex \cite{VaqueiroContreras2019}. So, it appears that the $\mathrm{B_{s}O_{2}}$
complex is a multi-stable defect with at least three experimentally
confirmed atomic configurations for the neutral charge state ($\textrm{X}^{0}$,
$\textrm{A}^{0}$, and $\textrm{A}'^{0}$/$\textrm{A}''^{0}$). Because
of the existence of a rather high energy barrier between the $\textrm{A}'^{0}$/$\textrm{A}''^{0}$
and $\textrm{A}^{0}$ configurations the metastable $\textrm{A}'^{0}$/$\textrm{A}''^{0}$
configurations are long-living at room temperature. Only at temperatures
above 150~$^{\circ}\textrm{C}$ the $\textrm{A}'/\textrm{A}''\rightarrow\textrm{A}/\textrm{D}$
reactions occur efficiently.

\subsection{Atomic hydrogen in silicon\label{subsec:hydrogen}}

\begin{figure}
\begin{centering}
\includegraphics[width=7.5cm]{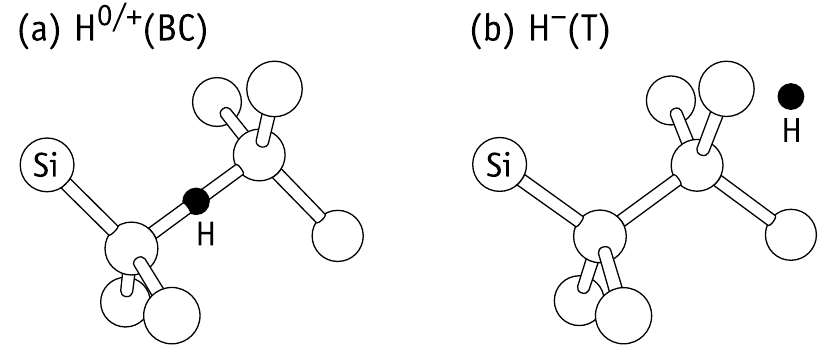}
\par\end{centering}
\caption{\label{FIG:HYDROGEN}Illustration of the minimum-energy atomic configurations
for hydrogen interstitials in silicon, as derived from first-principles
calculations (a) the positive charge and neutral charge states, and
(b) the negative charge state.}
\end{figure}

Hydrogen is a very common impurity in silicon. It is quite mobile
at room temperature and extremely reactive in its atomic form. As
a result, it readily combines with other impurities and itself, leaving
only a small fraction of isolated hydrogen atoms. These exist in three
charge states and exhibit negative-$U$ behaviour with the H-related
$(0/+)$ donor level lying above the $(-/0)$ acceptor level, where
neutral H$^{0}$ is a metastable species \cite{Herring2001}. \emph{Ab-initio}
calculations predict that both H$^{0}$ and H$^{+}$ have minimum
energy positions at the Si-Si bond centre (BC), while H$^{-}$ is
located at an interstitial tetrahedral (T) lattice site \cite{Walle1989}.
Representative structures are shown in Figure~\ref{FIG:HYDROGEN}.
The resultant negative-$U$ characteristic of isolated hydrogen is
technologically very significant, impacting on hydrogen’s ability
to passivate defects and its diffusivity. Several comprehensive reviews
have been published which include the negative-$U$ aspects of hydrogen
in silicon \cite{Pankove1991,Walle2006,Peaker2008,Estreicher2014}.

The experimental study of isolated hydrogen in silicon has proved
to be very difficult and it is only by the application of multiple
techniques and calculations that a good understanding has been achieved.
The fundamental problem is that hydrogen is both highly reactive and
very mobile. The end result has been that some experiments claiming
to characterise isolated hydrogen actually derive from hydrogen complexes,
commonly H-O and H-C. A methodology to avoid this, which has met with
success, is to implant protons at low temperature and undertake measurements
at low temperature. The technique was first used by Stein \cite{Stein1975}
who showed unambiguously that the line at 1998~cm$^{-1}$ seen in
LVM absorption studies was related to hydrogen at the bond centre
position. Low temperature implantation at $T=45$~K was used by Holm
\cite{Holm1991} to study n-type Si using DLTS, and an $E_{\textrm{c}}-E(0/+)=0.16$~eV
donor level ($\textrm{E}3'$) was observed, being later ascribed to
bond-centred H(BC) . This was followed by detailed studies by Bonde-Nielsen
\emph{et~al.} \cite{Nielsen2002} on hydrogen near the T site, which
quantified the impact of oxygen on nearby sites and determined the
energy of the hydrogen acceptor to be $E_{\textrm{c}}-E(-/0)=0.68$~eV.

\subsection{DX centres in III-V alloys\label{subsec:dx}}

DX centres are a special case of negative-$U$ systems in which the
addition of dopants, expected to form shallow donors in III-V alloys,
gives rise to persistent photoconductivity and deep acceptors. The
phenomenon was first reported in detail by Craford \emph{et~al.}
in 1968 \cite{Craford1968} describing Hall effect measurements of
$\textrm{GaAs}_{1-x}\textrm{P}_{x}$ with $x\approx0.3$ doped with
Te or S. The phenomenon was erroneously believed to be the result
of a donor `$D$' reacting with an unknown defect `X' to produce
a deep state (hence `$DX$'). This was a result of studies of $\textrm{Al}_{x}\textrm{Ga}_{1-x}\textrm{As}$
with $x>0.22$ doped with Si, Ge, Sn, S, Se or Te using Hall effect,
photo-capacitance and DLTS (see for example Refs.~\cite{Henry1977,Lang1979}).
The story of the $DX$ centres is a classic research history spanning
well over 25 years starting with a technological driver of achieving
adequately high doping in III-V heterojunction lasers and transistors
and involving some of the world’s leading semiconductor groups combining
theory and experiment. After many plausible explanations were explored
it is now evident that the $DX$ centre is a classic example of negative-$U$
behaviour. The saga up to 1990 has been reviewed by Mooney \cite{Mooney1990}
and some key issues are covered in the introductory section of the
current paper.

\begin{figure}
\begin{centering}
\includegraphics[width=7.5cm]{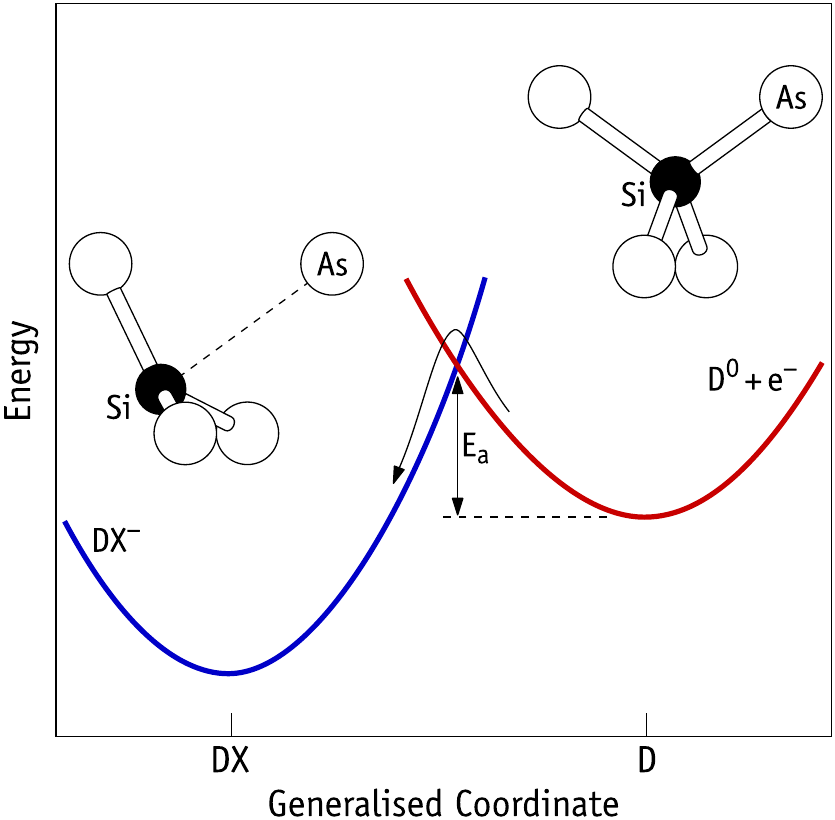}
\par\end{centering}
\caption{\label{FIG:DX}Configuration coordinate diagram of the Si $DX$ centre
in $\textrm{Al}_{x}\textrm{Ga}_{1-x}\textrm{As}$. In the $DX$ configuration
(left), the Si atom is displaced from its substitutional site. In
the shallow-donor configuration (right), the Si atom occupies the
substitutional site.}
\end{figure}

The most extensively studied $DX$ centre is that of Si in $\textrm{Al}_{x}\textrm{Ga}_{1-x}\textrm{As}$.
$DX$ behaviour is observed for the case where $x>0.22$ or at lower
aluminium concentrations under hydrostatic pressure, for example $DX$
behaviour of Si in GaAs is observed at pressures $P>2$~GPa. In the
$DX$ configuration the silicon accepts an electron and is negatively
charged. The charge state has been determined experimentally by Dobaczewski
et~al. \cite{Dobaczewski1991} from direct measurement of the capture
cross section. As shown in Figure~\ref{FIG:DX}, this change of charge
state results in the relaxation of the Si from its substitutional
position on the group III site, to an interstitial site neighbouring
three As atoms. In this siting, the Si does not act as a conventional
thermally ionised donor and does not contribute to the conductivity.

If the system is exposed to light of a sufficiently high energy, electrons
are excited from the Si to the conduction band and the $DX$ centre
reverts to the substitutional site where it is a metastable neutral
donor. At temperatures $T<180$~K the electron does not have enough
energy to surmount the barrier ($E_{\textrm{a}}\sim0.2$~eV) shown
by the arrow in Fig.~\ref{FIG:DX}, and so the additional conductivity
resulting from the shallow donor state induced by the light remains
for hours or at lower temperatures for days or even weeks. This bizarre
behaviour is referred to as persistent photoconductivity and is a
signature of $DX$ behaviour.

The identification of the substitutional-interstitial motion evolved
from a sequence of calculations and experiments discussed in Mooney’s
review \cite{Mooney1990}. However, a series of revealing experiments
published subsequently provided unambiguous detail of this behaviour
\cite{Dobaczewski1992b,Dobaczewski1995}. In this work, high resolution
Laplace DLTS was analysed to provide detail of the influence of the
local environment on the electron thermal emission process from $DX$
centres in a range of compositions of $\textrm{Al}_{x}\textrm{Ga}_{1-x}\textrm{As}$.
The direct comparison of this process for centres related to Si which
can replace gallium or aluminium, with that observed for a group-VI
donor (tellurium), which resides in the arsenic sublattice, enabled
the configuration of atoms to be determined when the centre is in
the ground state. This shows that the $DX$(Si) defect ionisation
process is associated with interstitial-substitutional motion of the
silicon atom. However, in the case of the $DX$(Te) centre, the spectra
shows peaks from two groups which are attributed to interstitial substitutional
motion of the neighbouring aluminium or gallium atoms, respectively.

$DX$ behaviour has been observed in many III-V and II-VI semiconductors,
and recently, persistent photoconductivity has been reported in oxides
(\emph{e.g.} ZnO and SrTiO$_{3}$), which has been interpreted as
$DX$ behaviour associated with negative-$U$ characteristics \cite{Bhatt2019,Tarun2013}.

\subsection{The Ga-$\textrm{O}_{\textrm{As}}$-Ga defect in GaAs\label{subsec:ga-o-ga}}

The Ga-$\textrm{O}_{\textrm{As}}$-Ga defect in GaAs is often referred
to as substitutional oxygen (on the arsenic site) or arsenic-vacancy-oxygen
($V_{\textrm{As}}$-O) complex \cite{Stavola2018}. It shows up as
a group of three local vibrational mode (LVM) absorption bands with
identical fine structure. The bands, generally referred to as $\textrm{A}$,
$\textrm{B}'$ and $\textrm{B}$, appear at 731~cm$^{-1}$, 714~cm$^{-1}$
and 715~cm$^{-1}$. Band $\textrm{B}'$ emerges during the intermediate
stages of optical-induced conversion from $\textrm{A}$ to $\textrm{B}$
using light with $h\nu>0.8$~eV (below band-gap light). This was
interpreted as an electron-transfer process from EL2 to the conduction
band, and subsequent capture by the Ga-$\textrm{O}_{\textrm{As}}$-Ga
complex. EL2 is an As antisite-related defect, responsible for a pinning
$E(0/+)$ level, effectively working as a reservoir of electrons at
$E_{\textrm{c}}-0.75$~eV \cite{Look1999}. Hence, $\textrm{A}$,
$\textrm{B}'$ and $\textrm{B}$ bands were ascribed to LVMs of the
same complex in charge states $q$, $q-1$ and $q-2$, respectively
\cite{Alt1990}.

The $\textrm{A}$ and $\textrm{B}$ bands were first reported in semi-insulating
material and assigned to EL2 by Song $et\,al.$ \cite{Song1987}.
However, this connection could not be correct and was quickly supplanted
by a model involving one O atom with two equivalent Ga-O bonds \cite{Zhong1988}.
That way, the triple-peak fine structure with intensity ratio of about
2:3:1 on each band (from high to low frequency peaks), could be explained
by a combination of natural occurring Ga isotopes. Considering that
$^{69}$Ga and $^{71}$Ga are found in a proportion of about 3:2,
a natural population of $^{69}$Ga-X-$^{69}$Ga, $^{69}$Ga-X-$^{71}$Ga,
and $^{71}$Ga-X-$^{71}$Ga units corresponds to an intensity ratio
of $9:(2\times6):4=2.25:3:1$, respectively \cite{Zhong1988,Schneider1989}.
Based on sample history arguments and semi-empirical calculations,
the X element was suggested to be oxygen \cite{Zhong1988}. This was
confirmed shortly after with the observation of a $^{18}$O-isotope
shift of the bands by Schneider and co-workers \cite{Schneider1989}.

Besides possessing a Ga-O-Ga unit with two equivalent Ga-O bonds,
we know from stress-splitting FTIR absorption that the $\textrm{A}$
band involves a $\langle110\rangle$-oriented vibration dipole \cite{Song1990}
and the defect should have $C_{2v}$ symmetry. These experiments also
concluded that Ga-$\textrm{O}_{\textrm{As}}$-Ga does not reorient
like VO in Si.

\begin{figure}
\begin{centering}
\includegraphics[width=7cm]{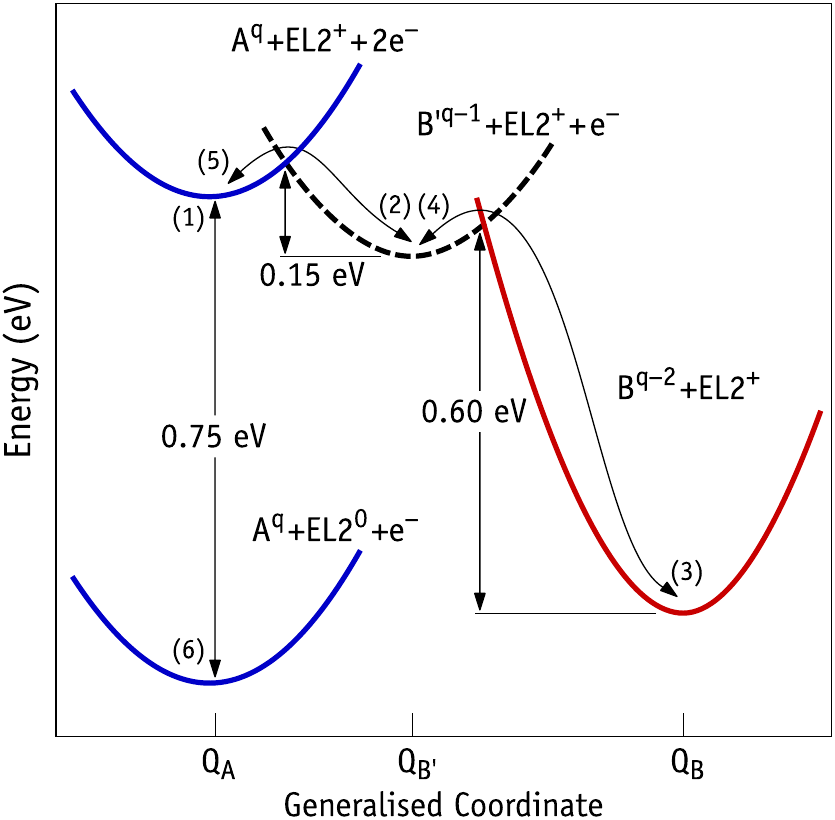}
\par\end{centering}
\caption{\label{FIG:Ga-O-Ga}Configuration coordinate diagram for the Ga-O-Ga
defect in GaAs based on experimental observations. Arrow heads indicate
electron capture and emission processes described in the text.}
\end{figure}

The negative-$U$ nature of the Ga-$\textrm{O}_{\textrm{As}}$-Ga
defect was found by Alt, who performed a series of enlightening experiments,
combining Fourier-transform infra-red (FTIR) absorption, photo-excitation
using monochromatic light with energy spanning the band gap and annealing
\cite{Alt1990,Alt1991}. The configuration coordinate diagram presented
in Figure~\ref{FIG:Ga-O-Ga} was constructed from the experimental
data. Upon cooling the sample from room temperature to 10~K in dark,
only band $\textrm{A}$ was observed, indicating that this should
be the ground state in semi-insulating GaAs (for $E_{\textrm{F}}=E_{\textrm{c}}-0.75$~eV).
State A is here assumed to have a reference charge state $q$, thus
being referred to as $\textrm{A}^{q}$.

After applying a short pulse of light with $0.8\,\textrm{eV}<h\nu<1.5\,\textrm{eV}$,
band $\textrm{A}$ was partially quenched and that corresponded to
a proportional growth of band $\textrm{B}'$. As referred to above,
this was assigned to the photoionisation of EL2 followed by electron
capture by Ga-$\textrm{O}_{\textrm{As}}$-Ga. Importantly, band $\textrm{B}$
was not visible at this point. This process is indicated in Figure~\ref{FIG:Ga-O-Ga}
by the arrow heads with labels (1) and (2). The $\textrm{B}'$ band
was therefore assigned to the $q-1$ charge state ($\textrm{B}'^{q-1}$).

Applying a 95~K dark annealing treatment to samples where only $\textrm{A}$
and $\textrm{B}'$ bands were present, led to the quenching of $\textrm{B}'$
and to the concurrent growth of bands $\textrm{A}$ and $\textrm{B}$.
The argument was that by warming the sample at 95~K, electrons bound
to $\textrm{B}'^{q-1}$ were thermally promoted into the conduction
band, thus forming $\textrm{A}^{q}$ again (process with arrow head
(5) in Figure~\ref{FIG:Ga-O-Ga}). The relatively small capture cross
section of ionized EL2, also allowed free electrons to be captured
by other $\textrm{B}'^{q-1}$ defects, thus leading to the growth
of band $\textrm{B}$. This was ascribed to the formation of a $q-2$
charge state, namely $\textrm{B}{}^{q-2}$ (process number (3) in
Figure~\ref{FIG:Ga-O-Ga}). The activation energy for this conversion
was measured as $E_{\textrm{a}}=0.15$~eV, and was assigned to the
thermal emission barrier of $\textrm{B}'^{q-1}\rightarrow\textrm{A}^{q}+\textrm{e}^{-}$
\cite{Alt1990,Skowronski1990}. As pointed out by Alt \cite{Alt1990},
process (3) and (5) consisted on a disproportionation of metastable
$\textrm{B}'^{q-1}$ with negative-$U$ ordering of levels. It was
also noted that full $\textrm{A}\rightarrow\textrm{B}$ band conversion
could be achieved by subjecting the samples for long illumination
times with $0.8\,\textrm{eV}<h\nu<1.5\,\textrm{eV}$ at $T<70$~K.
At these temperatures, the thermal emission channel (5) in Figure~\ref{FIG:Ga-O-Ga}
was suppressed and all defects ended up in state $\textrm{B}{}^{q-2}$.

At higher temperatures ($T\geq180$~K), band $\textrm{B}$ starts
to decrease in intensity at a rate identical to the recovery of band
$\textrm{A}$. Full recovery of band $\textrm{A}$ is achieved when
$\textrm{B}$ vanishes. This was suggested to result from a two-electron
emission sequence $\textrm{B}{}^{q-2}\rightarrow\textrm{B}'{}^{q-1}+\textrm{e}^{-}\rightarrow\textrm{A}^{q}+2\textrm{e}^{-}$,
where the limiting step is the first one (with larger barrier). The
above emission sequence is represented in Figure~\ref{FIG:Ga-O-Ga}
as steps with arrow heads (4) and (5), respectively. At these temperatures,
the thermally emitted electrons were shown to be captured and trapped
at the deeeper EL2 state (step (6) of Figure~\ref{FIG:Ga-O-Ga}).
The activation energy for the $\textrm{B}\rightarrow\textrm{A}$ conversion
was measured as $E_{\textrm{a}}=0.60$~eV, thus suggesting that the
binding energy of the second electron captured by the Ga-$\textrm{O}_{\textrm{As}}$-Ga
defect, exceeded the binding energy first one (0.15~eV).

We note that a more cautious analysis of the above picture would include
the capture barriers for steps (2) and (3). DLTS measurements give
activation energies of 0.58 eV and 0.14 eV for electron emission from
$\textrm{B}^{q-2}$ and $\textrm{B}'^{q-1}$, respectively \cite{Kaufmann1991}.
As far as we are aware, capture barriers were not measured. However,
judging from Hall effect measurements, which find a disproportionating
$(q\!-\!2/q)$ transition at $E_{\textrm{c}}-0.43$~eV, and considering
the above emission barriers, the sum of first and second capture barriers
should be in the meV range \cite{Alt1989a,Alt1991}. Also interestingly,
based on its DLTS fingerprint, electron emission from $\textrm{B}^{q-2}$
was assigned to EL3, a previously known deep trap with unknown composition
\cite{Kaufmann1991}.

The first atomistic model claiming to account for the above data was
a $\langle001\rangle$-displaced substitutional oxygen atom at the
arsenic site ($\textrm{O}_{\textrm{As}}$) \cite{Schneider1989},
resembling the vacancy-oxygen complex in silicon. $\textrm{O}_{\textrm{As}}$
possesses a Ga-O-Ga unit aligned along the $\langle110\rangle$ direction,
thus being compliant with the conclusions from the polarised absorption
experiments \cite{Song1990}. Jones and Öberg \cite{Jones1992} inspected
the $\mathrm{O_{As}}$ defect in H-terminated GaAs spherical clusters
using local density functional theory. The paramagnetic $\textrm{O}_{\textrm{As}}^{=}$
state was found to be metastable and was ascribed to band $\textrm{B}'$.
Bands $\textrm{A}$ and $\textrm{B}$ were therefore connected to
diamagnetic states with $\textrm{O}_{\textrm{As}}^{-}$ and $\textrm{O}_{\textrm{As}}^{-3}$,
respectively. While the calculations of Ref.~\cite{Jones1992} accounted
for the experimentally observed red-shift of the oxygen vibrational
frequency of $\textrm{B}$ with respect to $\textrm{A}$, this result
could not be reproduced by subsequent calculations \cite{Mattila1996,Taguchi1998,Pesola1999}.
Additionally, and perhaps more problematic for the $\textrm{O}_{\textrm{As}}$
model, is that it cannot explain the lack of reorientation of the
defect under the effect of uniaxial stress (although this was speculated
to be due to a nearly isotropic stress tensor \cite{Song1990}).

Alternatively, Pesola \emph{et~al.} \cite{Pesola1999} proposed a
defect model in which the two Ga radicals (not connected to oxygen)
in the off-site $\textrm{O}_{\textrm{As}}$ defect are replaced by
As. This complex was referred to as $(\mathrm{{As}_{Ga}})_{2}\textrm{-}\mathrm{{O}_{As}}$
and incorporates a Ga-O-Ga unit like in $\textrm{O}_{\textrm{As}}$.
The first-principles calculations of Ref.~\cite{Pesola1999} indicated
that $(\mathrm{{As}_{Ga}})_{2}\textrm{-}\mathrm{{O}_{As}}$ shows
a negative-$U$ ordering of $E(0/+)$ and $E(-/0)$ transitions, and
the calculated O-LVM frequencies of charge states $+$, $0$ and $-$
closely follow the observed frequency variations of bands $A$, $\textrm{B}'$
and $\textrm{B}$, respectively. The model naturally accounts for
the lack of reorientation under uniaxial stress. However, it clashes
with optically detected electron-nuclear double resonance (OD-ENDOR),
which reported that as much as 60\% of spin-density of the $\textrm{B}'^{0}$
state (with spin-1/2) is localized on the first shell of Ga atoms
\cite{Koschnick1997} -- Figure~2 of Ref.~\cite{Pesola1999} clearly
shows that the $(\mathrm{{As}_{Ga}})_{2}\textrm{-}\mathrm{{O}_{As}}$
model has most of the density on As shells and virtually none on Ga
shells.

The origin for the negative-$U$ ordering of levels in the Ga-$\textrm{O}_{\textrm{As}}$-Ga
complex is not clear. The simpler $\textrm{O}_{\textrm{As}}$ substitution,
although incompatible with the vibrational absorption data, could
explain the effect based on strong charge-dependent Ga-Ga reconstructions.
This feature was found theoretically for the As vacancy ($V_{\textrm{As}}$)
in GaAs \cite{Mellouhi2005}, and used as a justification for $V_{\textrm{As}}^{0}$
and $V_{\textrm{As}}^{=}$ showing a negative-$U$. On the other hand,
the $(\mathrm{{As}_{Ga}})_{2}\textrm{-}\mathrm{{O}_{As}}$ model has
a mid-gap state localised on two As radicals, which were suggested
to induce strong forces on neighbouring atoms as a function of occupancy
\cite{Pesola1999}. This effect does not find a parallel in the (positive-$U$)
Ga vacancy ($V_{\textrm{As}}$), whose As radicals induce weak deformations
\cite{Mellouhi2005}.

Recently, the $\textrm{O}_{\textrm{As}}$ defect was investigated
using modern hybrid density functional theory \cite{Colleoni2014}.
$\textrm{O}_{\textrm{As}}^{+}$ and $\textrm{O}_{\textrm{As}}^{0}$
were found to be most stable at the perfect tetrahedral site. The
highest occupied state of $\textrm{O}_{\textrm{As}}^{0}$, which is
paramagnetic, is an s-like orbital on the oxygen atom, and therefore
not compliant with the OD-ENDOR data. $\textrm{O}_{\textrm{As}}^{-}$
had an off-site Ga-$\textrm{O}_{\textrm{As}}$-Ga configuration with
a localised acceptor state on a reconstructed Ga-Ga pair. The $E(0/+)$
and $E(-/0)$ levels were calculated at $E_{\textrm{c}}-0.71$~eV
and $E_{\mathrm{c}}-0.54$~eV, thus showing a positive-$U$ ordering.
Clearly the Ga-$\textrm{O}_{\textrm{As}}$-Ga defect should be revisited
by both theory and experiments.

\subsection{Electrical levels of selected negative-$U$ defects in semiconductors\label{subsec:tab-negu}}

Table~\ref{TAB:NEGU} lists the electronic characteristics of a selection
of negative-$U$ defects in semiconductors. It should be noted that
the table is not comprehensive. We only considered defects for which
the negative-$U$ behaviour is confirmed by solid experimental results,
where their electronic properties are relatively well understood and
the atomic structures in different charge states have been predicted
by experiments and \emph{ab-initio} modelling. Negative-$U$ properties
have been suggested for a number of other defects in semiconductor
materials, \emph{e.g.}, for interstitial boron-interstitial oxygen
complex and hydrogen-related shallow donors in silicon \cite{Makarenko2014,Mukashev2000}.
Besides silicon, atomic hydrogen has also been claimed to show negative-$U$
behaviour in many other semiconductors \cite{VandeWalle2003,VandeWalle2006}.
Further examples could be enumerated. However, no solid experimental
evidence confirming these suggestions has been presented yet. Perhaps
it is worth mentioning the interesting case of the Si trivacancy,
which has been briefly discussed in Section~\ref{subsubsec:barriers}.
It has been argued in Refs.~\cite{Markevich2009,Coutinho2012} that
$V_{3}$ in Si can be considered as a defect with negative-$U$ properties
as it has the $E(=/-)$ acceptor level at $E_{\textrm{c}}-0.21$~eV
below the $E(-/0)$ level at $E_{\textrm{c}}-0.07$~eV. However,
the levels mentioned above are energy levels of $V_{3}$ in different
configurations (part of the hexagonal ring and four-fold coordinated
shown in Figures~\ref{FIG:V3-REORIENT}(a) and \ref{FIG:V3-REORIENT}(d),
respectively), which are separated by rather high energy barriers.
Because of these barriers the paired emission or capture of electrons
(a signature of negative-$U$ defects) has not been observed for the
$V_{3}$ defect, and therefore, it has not been included into the
table.

The transition levels of Table~\ref{TAB:NEGU} are cast as $\mbox{\ensuremath{E(q\,/\,q\!+\!1)}}$
and $E(q\!\!-\!\!1\,/\,q)$, where $q$ is a reference charge state.
For negative-$U$ defects $E(q\,/\,q\!\!+\!\!1)>E(q\!\!-\!\!1\,/\,q)$,
and the $q$ charge state corresponds to the metastable state which,
under equilibrium conditions, disproportionates into stable $q+1$
and $q-1$ charge states.

\cleardoublepage{}

\noindent 
\begin{sidewaystable}
\caption{\label{TAB:NEGU}Electrical levels of selected negative-$U$ defects
in semiconductors and respective key references. All energies in eV.
Starred ({*}) defects are discussed in detail either below or elsewhere
in this paper.}

\centering{}%
\begin{tabular}{llclcl>{\raggedright}p{6cm}l}
\toprule 
Host & Defect & $q/q+1$ & $E(q/q+1)$ & $q-1/q$ & $E(q-1/q)$ & Comment & Key references\tabularnewline
\midrule 
Si & Isolated vacancy ({*}) & $+/+\!+$ & $E_{\text{v}}+0.13$ & $0/+$ & $E_{\text{v}}+0.05$ & DLTS, EPR, p-type, low temperature irradiation & \cite{Baraff1979,Watkins1980}\tabularnewline
Si & Interstitial B ({*}) & $0/+$ & $E_{\text{c}}-0.13$ & $-/0$ & $E_{\text{c}}+0.3$7 & DLTS, ODLTS, EPR, p- and n-type, n+p \& p+n diodes, low temperature
irradiation & \cite{Troxell1980,Harris1982,Harris1987}\tabularnewline
Si & Hydrogen-Oxygen & $0/+$ & $E_{\textrm{c}}-0.16$ & $-/0$ & $E_{\textrm{c}}-0.79$ & DLTS, n-type Cz, low temperature proton implantation & \cite{Nielsen2002}\tabularnewline
Si & Hydrogen ({*}) & $0/+$ & $E_{\textrm{c}}-0.16$ & $-/0$ & $E_{\textrm{c}}-0.68$ & C transients, PH dissociation, EPR, n-type, low temperature proton
implantation & \cite{Herring2001}\tabularnewline
Si & $\textrm{B}_{\textrm{s}}\textrm{O}_{2}$ ({*}) & $0/+$ & $E_{\textrm{v}}+0.56$ & $-/0$ & $E_{\textrm{v}}+0.04$ & DLTS, LDLTS, admittance spectroscopy, p type CZ, heat-treatments at
400-500~$^{\circ}$C & \cite{VaqueiroContreras2019,Markevich2019}\tabularnewline
Si & $\mathrm{C_{i}O_{i}H}$ & $0/+$ & $E_{\textrm{c}}-0.043$ & $-/0$ & $E_{\textrm{c}}-0.11$ & Hall effect, DLTS, FT-IR, n-type Cz, heat-treatments 350~$^{\circ}$C
electron irradiated hydrogenated samples & \cite{Markevich1994,Markevich1997,Coutinho2001}\tabularnewline
Si & Self-interstitial ($I$) & $+/+\!+$ & $E_{\textrm{c}}-0.39$ & $0/+$ & ? & DLTS, EPR, p-type, low temperature implantation with alpha particles & \cite{Gorelkinskii2009}\tabularnewline
Si & $I\textrm{O}_{2}$ ({*}) & $+/+\!+$ & $E_{\textrm{v}}+0.36$ & $0/+$ & $E_{\textrm{v}}+0.12$ & Hall effect, DLTS, FT-IR, carbon lean p-type Cz, room temperature
electron irradiation & \cite{Lindstrom2001,Markevich2005a,Markevich2005b}\tabularnewline
\midrule 
Si & O-related TDDs & $+/+\!+$ & $E_{\textrm{c}}-0.16$ & $0/+$ & TDD-0, $E_{\textrm{c}}-0.75$ & \multirow{3}{6cm}{Hall effect, photoconductivity, FT-IR, n-type Cz, heat treatments
400-500~$^{\circ}$C} & \multirow{3}{*}{\cite{Makarenko1985,Makarenko1988,Latushko1986}}\tabularnewline
 &  &  &  & $0/+$ & TDD-1, $E_{\textrm{c}}-0.48$ &  & \tabularnewline
 &  &  &  & $0/+$ & TDD-2, $E_{\textrm{c}}-0.29$ &  & \tabularnewline
\midrule 
Ge & O-related TDDs & $+/+\!+$ & $E_{\textrm{c}}-0.04$ & $0/+$ & TDD-0, $E_{\textrm{c}}-0.48$ & \multirow{3}{6cm}{Hall effect, photoconductivity, FT-IR, n type Cz, heat treatments
300-400~$^{\circ}$C} & \multirow{3}{*}{\cite{Litvinov1988,Clauws1991}}\tabularnewline
 &  &  &  & $0/+$ & TDD-1, $E_{\textrm{c}}-0.34$ &  & \tabularnewline
 &  &  &  & $0/+$ & TDD-2, $E_{\textrm{c}}-0.26$ &  & \tabularnewline
\midrule 
4H-SiC & C vacancy ({*}) & $-/0$ & $E_{\textrm{c}}-0.42$ & $=/-$ & $E_{\textrm{c}}-0.64$ & DLTS, Laplace-DLTS, EPR, n-type, traces in all as-grown samples & \cite{Hemmingsson1998,Son2012,Capan2018}\tabularnewline
6H-SiC & C vacancy ({*}) & $-/0$ & $E_{\textrm{c}}-0.20$ & $=/-$ & $E_{\textrm{c}}-0.43$ & DLTS, Laplace-DLTS, n-type, traces in all as-grown samples & \cite{Hemmingsson1999,Koizumi2013}\tabularnewline
InP & M-center & $+/+\!+$ & $E_{\textrm{c}}-0.18$ & $0/+$ & $E_{\textrm{c}}-0.42$ & TSCAP, DLTS, optical ionisation, irradiated samples & \cite{Levinson1983a,Levinson1983b,Stavola1984,Wager1985}\tabularnewline
GaAs & Ga-$\textrm{O}_{\textrm{As}}$-Ga ({*}) & $0/+$ & $E_{\textrm{c}}-0.14$ & $-/0$ & $E_{\textrm{c}}-0.58$ & DLTS, optical ionisation & \cite{Alt1990,Kaufmann1991,Pesola1999}\tabularnewline
GaAsP & $DX$ centre & \multicolumn{4}{l}{See text, properties depend on alloy composition} & Hall effect, optical absorption & \cite{Craford1968}\tabularnewline
AlGaAs & $DX$ centre & \multicolumn{4}{l}{See text, properties depend on alloy composition} & DLTS, photo-capacitance, Hall effect & \cite{Lang1977,Lang1979,Chadi1988,Dobaczewski1991,Baraff1992}\tabularnewline
ZnO & Oxygen vacancy & $+/+\!+$ & $E_{\textrm{c}}-0.4$ & $0/+$ & $E_{\textrm{c}}-1.6$ & DLTS, theory & \cite{Alkauskas2011,Hupfer2016}\tabularnewline
GaSb & $DX$(S) & $0/+$ & $E_{\textrm{c}}-0.22$ & $-/0$ & $E_{\textrm{c}}-0.25$ & DLTS & \cite{Dobaczewski1991}\tabularnewline
\bottomrule
\end{tabular}
\end{sidewaystable}

\clearpage{}

\section{Conclusions\label{sec:conclusions}}

The present review begins with an historical perspective of the topic
of `negative-$U$ defects in semiconductors'. We start from the
early 1980’s, where the first reports of this class of defects emerged
in the literature. The account includes several difficulties identified
at the time, mostly due to meta-stability, or in some cases multi-stability,
of one of the three charge states involved. We provide the reader
with an introduction to the essential physics of the workings of negative-$U$
defects (electronic correlation, electron-phonon coupling, disproportionation
and defect transition levels). We present the state-of-the-art regarding
the experimental and theoretical methods that are currently used for
the characterisation of negative-$U$ defects. These subjects are
presented in detail and they are accompanied with several successful
studies from the literature. This includes experimental methods based
either on defect thermodynamics or defect kinetics (monitored by several
techniques, like Hall effect, DLTS or optical absorption), as well
as numerical methods that can help us in the construction of detailed
configuration coordinate diagrams (with the calculation of transition
levels, transition rates and transformation barriers). To wrap up,
we showcase a few experimental and theoretical analyses of some of
the most impacting negative-$U$ defects in group-IV and compound
semiconductors, leaving a summary of the characteristics of many others,
along with the most relevant references in a tabular format.

A negative-$U$ defect has a metastable state with $N$ electrons
that disproportionate into $N-1$ and $N+1$ species under equilibrium
conditions. This is only possible in many-atom systems, and thanks
to the conversion of electron correlation energy into bond formation
energy, either during ionisation or capture of carriers. In these
defects, the stable $N-1$ and $N+1$ states are normally diamagnetic
close-shell systems, making them unnoticeable by paramagnetic resonance
techniques, unless some sort of excitation is provided to the sample.
The metastability of the $N$-electron state works like a barrier
separating the $N-1$ and $N+1$ species. When one of the latter is
a shallow donor or a shallow acceptor, their occupation via deep-shallow
excitation induced by illumination, usually results in a concentration
increase of free carriers. This effect is often measurable as a persistent
photoconductivity, which depending on the temperature, may last for
several hours, until the defects surmounts a barrier associated with
a charge capture event followed by transformation and return to the
ground state configuration.

The above features make junction spectroscopy methods a primary choice
for the characterisation of negative-$U$ defects in semiconductors.
However, the complexity of their behaviour often requires the usage
of complementary techniques, as well as special tricks, like illumination
or thermal quenching, which allow researchers to play with thermodynamics
and kinetics, and by doing so, to isolate the defect in one of its
three states, or to investigate the conversions between them. Defect
modelling has also been pivotal. Since the pioneering work on the
silicon vacancy, theories, numerical algorithms and hardware have
evolved dramatically. Density functional related methods are perhaps
the most successful cases. These have been able to provide us with
a detailed quantum mechanical description of the interactions between
trapped carriers and the many-body cloud of electrons comprising the
defective semiconductor.

Finally, we would like to emphasise the following concluding remark
-- we have achieved a rather deep understanding of the electronic
and atomistic structure of many negative-$U$ defects in semiconductors.
For sure, that was achieved thanks to concurrent and combined experimental
and theoretical work.

\ack{}{}

JC thanks the support of the i3N project, Refs. UIDB/50025/2020 and
UIDP/50025/2020, financed by the FCT/MEC in Portugal. VM and ARP have
been supported by the UK EPSRC for this work under projects EP/M024911/1
and EP/P015581/1

\section*{ORCID iDs}
\begin{lyxlist}{00.00.0000}
\item [{J~Coutinho}] \href{https://orcid.org/0000-0003-0280-366X}{https://orcid.org/0000-0003-0280-366X}
\item [{V~P~Markevich}] \href{https://orcid.org/0000-0002-2503-6144}{https://orcid.org/0000-0002-2503-6144}
\item [{A~R~Peaker}] \href{https://orcid.org/0000-0001-7667-4624}{https://orcid.org/0000-0001-7667-4624}
\end{lyxlist}
All authors contributed equally to this work.

\section*{References}{}

\bibliographystyle{iopart-num}
\phantomsection\addcontentsline{toc}{section}{\refname}

\providecommand{\newblock}{}

\end{document}